\documentclass[trackchanges,twocolumn]{aastex7}
\usepackage{natbib}
\usepackage[table]{xcolor}
\usepackage{framed}
\usepackage{multirow}
\usepackage{longtable}
\usepackage{siunitx}
\usepackage{booktabs}
\usepackage{caption}
\usepackage[T1]{fontenc}
\usepackage{hyperref}
\usepackage{newtxtext,newtxmath}
\usepackage{amsmath}

\colorlet{shadecolor}{blue!20}

\begin{document}

\title{Observing Co-Located Neutral and Ionized Gas-Phase Iron Depletion in the Magellanic Clouds}

\author[0000-0003-1147-831X]{Yun Qi Li}
\email{billyli@uw.edu}
\affiliation{Department of Astronomy, University of Washington, Seattle, WA 98195, USA}

\author[0000-0002-0355-0134]{Jessica K. Werk}
\email{jwerk@uw.edu}
\affiliation{Department of Astronomy, University of Washington, Seattle, WA 98195, USA}

\author[0000-0001-9200-169X]{Caleb R. Choban}
\email{cchoban@iu.edu}
\affiliation{Department of Astronomy, Indiana University, Bloomington, IN 47405, USA}

\author[0000-0001-6326-7069]{Julia Roman-Duval}
\email{duval@stsci.edu}
\affiliation{Space Telescope Science Institute, 3700 San Martin Drive, Baltimore, MD 21218, USA}

\author[0000-0003-0789-9939]{Kirill Tchernyshyov}
\email{ktcherny@gmail.com}
\affiliation{Department of Astronomy, University of Washington, Seattle, WA 98195, USA}

\author[0000-0002-7738-6875]{J. Xavier Prochaska}
\email{jxp@ucsc.edu}
\affiliation{Department of Astronomy \& Astrophysics, University of California Santa Cruz, 1156 High Street, Santa Cruz, CA 95064, USA}

\author[0000-0001-9654-5889]{Doyeon A. Kim}
\email{dakim@stsci.edu}
\affiliation{Space Telescope Science Institute, 3700 San Martin Drive, Baltimore, MD 21218, USA}

\author[0000-0002-7530-8857]{Arianna S. Long}
\email{aslong@uw.edu}
\affiliation{Department of Astronomy, University of Washington, Seattle, WA 98195, USA}

\begin{abstract}

Depletion is the observed phenomenon where gas-phase elemental abundances are reduced through accretion onto dust grains. We measure neutral gas-phase elemental abundances (S, Fe) in the Magellanic Clouds along 33 sightlines using high-resolution UV spectroscopy (HST/COS and HST/STIS), and compare them to ionized gas-phase abundances (S, Fe) adopted from the literature for six co-located H\,\textsc{ii} regions (with the furthest separation of $\lesssim3'$, 50 pc). Comparing S abundances show that S is minimally depleted in the H\,\textsc{ii} regions and surrounding diffuse ISM. However, we find that the gas-phase Fe abundances in H\,\textsc{ii} regions can be lower than those of the neighboring neutral ISM by 0.3 to 2 dex. This difference is likely an offset in the amount of Fe depleted into dust grains. As accretion of gas-phase Fe is likely not effective at the temperatures of the H\,\textsc{ii} regions, Fe depletion into solid form would have occurred in the dense atomic or molecular clouds prior to star formation. Stronger depletion in the H\,\textsc{ii} regions shows that Fe-bearing grains survive destruction in the first few million years following ionization. Our observations highlight that Fe depletion in H\,\textsc{ii} regions can be a useful tracer of Fe depletion in dense molecular clouds, which are challenging to observe directly via UV absorption. 

\end{abstract}

\section{Introduction} \label{introduction} 

Metals in the gas and solid phases make up only a small fraction of the total mass of the interstellar medium (ISM); however, it is this small portion ($\sim 1\%$ at solar metallicity) that determines the chemistry, energy, emissions, and evolution of the ISM \citep{draine_physics_2011}. A fraction of metals in the ISM reside within dust grains \citep{jenkins_unified_2009}, which play critical roles. Grains heat the diffuse ISM through the ejection of photoelectrons \citep{draine_physics_2011}. Grains serve as the sites that catalyze the formation of molecular hydrogen \citep{gould_interstellar_1963, hollenbach_surface_1971, draine_physics_2011}, and prevent photodissociation of molecules through shielding from FUV radiation \citep{draine_structure_1996}. Grains absorb stellar light in the ultraviolet (UV) and optical, and re-emit this energy in the far-infrared (FIR), which profoundly affects the observed spectral energy distribution (SED) of galaxies. Such emissions allow tracing the ISM of high redshift galaxies that are otherwise not observable in 21 cm emission \citep{galliano_interstellar_2018}. Ultimately, dust grains seed the formation of planetesimals and planets through complicated dynamical and chemical processes \citep[e.g.][]{drazkowska_planet_2023, birnstiel_dust_2024}. Metals in the gas and solid phase thus play dual and complementary roles in shaping the ISM and driving galaxy evolution. Studying them in concert is essential for quantifying and characterizing the processes that govern how galaxies form, grow, and change over cosmic time.

Metals are generally under-abundant in the gaseous ISM due to depletion, in which gas-phase elements become incorporated into solid grains. It is assumed that for a given galaxy, the total abundance of metals (relative to hydrogen) in stellar photospheres is equivalent to the combined gas-phase and solid-phase (depleted) metal content \citep[e.g.][]{jenkins_unified_2009, tchernyshyov_elemental_2015, jenkins_interstellar_2017, roman-duval_metal_2021, ritchey_distribution_2023, hamanowicz_metal-z_2024}. This assumption is first established by the crude correlation that elements which condense at higher minimum temperatures are observed to be more depleted \citep{field_interstellar_1974, spitzer_ultraviolet_1975, savage_interstellar_1996, jenkins_unified_2009}. Elemental depletions and their correlations with gas surface density have since been observationally characterized in the Milky Way \citep{jenkins_unified_2009, ritchey_distribution_2023}, the Magellanic Clouds \citep{tchernyshyov_elemental_2015, jenkins_interstellar_2017, roman-duval_metal_2019, roman-duval_metal_2021, roman-duval_metal_2022, roman-duval_metal_2022-1}, and other Local Group and nearby galaxies \citep{james_investigating_2014, james_tackling_2018, hamanowicz_metal-z_2024}. Directly supporting the connection between depletion and dust, \citet{decleir_first_2025} shows that Milky Way depletions of \citet{jenkins_unified_2009} and \citet{ritchey_distribution_2023} are positively correlated with the observed absorption from silicate grains. Gas-phase elemental abundances are primarily shaped by dust depletion, accounting for observed variations of up to $\sim$1--3 dex for refractory elements \citep{jenkins_unified_2009, tchernyshyov_elemental_2015, jenkins_interstellar_2017, roman-duval_metal_2019, hamanowicz_metal-z_2024}, while $\alpha$-element enhancement can be important in environments with recent star formation \citep{konstantopoulou_dust_2022, de_cia__2024}, elements such as oxygen (O), silicon (Si), and sulfur (S) are typically enhanced by $\sim$0.3 dex \citep{de_cia__2024}. 

Because direct dust observations alone cannot reveal elemental abundances, the depletion of metals provide a fundamental constraint to the composition of dust grains in current theoretical models \citep[e.g.][]{weingartner_dust_2001, hensley_astrodustpah_2023}. Given its refractory nature, the depletion of iron (Fe) is a particularly interesting test case. UV absorption studies, including \citet{jenkins_unified_2009} and subsequent work, show that $>90\%$ of Fe is depleted from the gas phase into solid grains in the Milky Way. Compared to the short grain destruction timescale by SNe shocks, it is suggested that the strong Fe depletion indicates dust grains regrow in the ISM via accretion of gas-phase species \citep{dwek_iron_2016}. The observed Fe depletion across galaxies of different intrinsic metallicities constrains the growth and destruction rates of Fe bearing grains, which encodes information of their physical properties such as size and charge \citep{zhukovska_evolution_2008, zhukovska_modeling_2016, zhukovska_iron_2018, choban_galactic_2022, choban_ashes_2026}. 

Elemental depletions in H\,\textsc{ii} regions have been studied via absorption and emission lines and offer a perspective on the presence of dust in ionized gas \citep[e.g.][]{howk_dust_1999}. Fe is strongly depleted within H\,\textsc{ii} regions \citep{rodriguez_fe_2005, izotov_chemical_2006, delgado-inglada_iron_2011, dominguez-guzman_homogeneity_2022}. A recent study by \citet{mendez-delgado_gas-phase_2024} show that the gas-phase Fe/O and Fe/N ratios in 452 local and extragalactic H\,\textsc{ii} regions anti-correlate with their metallicities. However, much is unclear regarding depletion in the H\,\textsc{ii} regions due to more complicated grain modification processes and uncertainties of gas-phase abundance measurements. Properties of dust in the H\,\textsc{ii} regions are strongly coupled to their spatial distribution \citep{faison_infrared_1998} and dynamics \citep{paladini_spitzer_2012}, while also depending heavily on the radiation field strength \citep{stephens_spitzer_2014} and morphology \citep{relano_dust_2016} of the regions themselves. Deriving accurate Fe depletion from emission lines proved challenging due to temperature estimation uncertainties \citep{peimbert_chemical_2003}, the discrepancy between collisionally excited lines (CELs) and recombination lines (RLs) \citep{toribio_san_cipriano_carbon_2017}, and ionization corrections \citep{rodriguez_fe_2005, mendez-delgado_gas-phase_2024}. Nevertheless, the depletion of Fe provides an explicit link between developments within the H\,\textsc{ii} region and the diffuse ISM scientific communities, offering a promising potential for constraining dust processes. 

Neutral-ionized comparisons of gas-phase elemental abundances have previously been performed primarily to address whether emission line abundances accurately represent the metallicity of a galaxy system \citep[e.g.][]{kunth_i_1986, aloisi_abundances_2003, lebouteiller_abundance_2004, james_tackling_2018}. Such comparisons are difficult to interpret in the context of dust depletion, due to low spacial resolution leading to averaging over abundance variations within the ISM. For example, in the low-metallicity galaxy I Zw 18, the ionized gas is observed to be metal-enriched compared to the neutral gas by a factor of 2 \citep{lebouteiller_chemical_2013, james_investigating_2014}, an effect attributed to either inhomogeneous mixing \citep{kunth_i_1986} or recent merger interactions \citep{lelli_dynamics_2012}. In the nuclear regions of M83, \citet{hernandez_first_2021} find that the neutral gas can have a higher metallicity than the ionized gas by $\sim$20\%, possibly because sulfur (S) measurements partially trace molecular gas in the nucleus, leading to an overestimate of S in the neutral phase. In the metal-poor blue compact dwarf galaxy NGC 5253, \citet{abril-melgarejo_mapping_2024} find that nitrogen (N), oxygen (O), and sulfur (S) are each more abundant in the ionized gas than the neutral gas by $\lesssim1$ dex, but Fe can be under-abundant in the ionized gas by less than $\sim0.1$ dex. In the other cases, neutral and ionized gas-phase metallicities show excellent agreement across galaxies spanning a wide range of metallicities \citep{james_tackling_2018, hernandez_first_2021, schady_comparing_2024, abril-melgarejo_mapping_2024, james_classy_2026}. Most recently, \citet{james_classy_2026} showed that Fe can be \textit{more} abundant in the neutral gas of star-forming galaxies than in the ionized gas by $0.25\pm0.47$ dex ($0.66\pm0.61$ dex after ionization corrections), suggesting either inhomogeneous mixing of Fe-rich material between ISM phases, or Fe depletion into dust in the ionized ejecta of Type II supernovae. 

In this study, we aim to compare the gas-phase depletion of Fe between the neutral and the ionized ISM to constrain dust grain processes. Our approach differs from previous comparisons by narrowing down the comparison to individual H\,\textsc{ii} regions and neutral gas measurements grouped within a parsec-scale. Within such distances, Fe depletion is unaffected by neutral ISM variations of $\gtrsim1$ dex on $\sim100$ pc scales \citep{jenkins_unified_2009, jenkins_interstellar_2017, roman-duval_metal_2021}, nor by variations of $\gtrsim0.5$ dex in separate H\,\textsc{ii} regions \citep{dominguez-guzman_homogeneity_2022}. The primary goal of this study is to constrain grain evolution processes. As our nearest galactic neighbors, the Magellanic Clouds offer several key advantages. They host numerous sightlines sufficiently bright for high-resolution neutral gas abundance analyses, and their H\,\textsc{ii} regions are bright enough that weaker emission lines (including those of Fe) can be detected and analyzed. Crucially, both the SMC and LMC are oriented nearly face-on, minimizing biases from metallicity gradients within the disk plane \citep{hernandez_first_2021} that complicate analogous Milky Way studies. The sightlines lie at physically comparable distances to the H\,\textsc{ii} regions, and their distinct radial velocities allow them to be cleanly separated from Milky Way foreground gas. Finally, both galaxies have been extensively observed over the past few decades, enabling us to draw on and reconcile independent sets of high-quality observations across the neutral and ionized phases.

For the neutral gas abundances, we measure absorption spectra gathered from the Mikulski Archive for Space Telescopes (MAST), consisting of programs detailed in Section~\ref{sightlines}. Our measurement procedure is developed in Section~\ref{methods}. Accumulation of H\,\textsc{ii} region elemental abundances are described in Section~\ref{hii-abundances}. We present results from the measurements and the comparison in Section~\ref{results}. The measurements are compared to existing studies in Section~\ref{compare-with-previous}. In Section~\ref{discussion}, we discuss the results in terms of dust grain evolution, and explore other factors such as ionization correction uncertainties, sightline blending, and metal-poor in-falls. 

\section{Data Selection and Measurements} \label{methods} 

Below, we describe the archival HST/COS and HST/STIS absorption spectra from several past programs that make up our primary dataset. We present our procedure for measuring the neutral gas column densities and compare them to existing measurements. Finally, we discuss the elemental abundances of the corresponding H\,\textsc{ii} regions, collected from the literature. 

\subsection{Archival UV Spectra and Neutral Gas Elemental Abundances} \label{hi-abundances}

The abundances of metals are typically represented in 12 + log$_{10}$(X/H), where X/H is the ratio of the measured gas-phase column density of the element to the measured gas-phase column density of hydrogen \citep[e.g.][]{jenkins_unified_2009, tchernyshyov_elemental_2015, jenkins_interstellar_2017, roman-duval_metal_2019}. The hydrogen column density consists of contributions from atomic and molecular hydrogen. In this section, we discuss our UV sightline selection, and then our measurements of both numerator and denominator in the above definition of metal abundance. 

\subsubsection{UV Sightline Selection} \label{sightlines}

Our primary goal with the archival UV spectra is to measure neutral gas-phase elemental abundances of sulfur (S) and iron (Fe). Other ions measured toward the H\,\textsc{ii} regions were considered; among the species for which 12 + log(X/H) are determined for the H\,\textsc{ii} regions, we note that nitrogen (N), oxygen (O), sulfur (S), and iron (Fe) abundances can also be determined in the neutral ISM through UV absorption spectroscopy. For the ISM of the Magellanic Clouds in particular, the N and O abundances are difficult to determine. N abundances are ambiguous due to saturation of N\,\textsc{i} transitions, previously noted by \citet{jenkins_interstellar_2017}. O abundances are ambiguous, even though they are detected via both strong and weak transitions in our spectra. For our sightlines, the strong O\,\textsc{i} $\lambda$1302 Å line is often significantly saturated. Conversely, the weak O\,\textsc{i} $\lambda$1355 Å line is seldom detected in our dataset. \citet{koenigsberger_hubble_2001} report that the O\,\textsc{i} $\lambda$1355 Å equivalent width toward SMC target HD 5980 is near the noise level of their data. \citet{de_cia__2024} comment that oxygen abundances derived from O\,\textsc{i} $\lambda$1355 Å are highly uncertain for both the SMC and the LMC ISM. For each detection of weak O\,\textsc{i} $\lambda$1355 Å absorption, we derived the O abundance and compared its depletion level to those of other elements. We find that O depletion differs significantly from other elements when using the F$_*$ method \citep{jenkins_unified_2009} with parameters from \citet{jenkins_interstellar_2017} and \citet{roman-duval_metal_2022}. The detected weak O absorption therefore likely suffers from a selection effect. Based on the above considerations, we did not further consider N and O; we choose S and Fe as the elements for which we measure 12 + log(X/H) in the neutral gas. Table~\ref{tab:lines-s-fe} lists the absorption lines we use to derive S and Fe abundances for each grating. 

\begin{table*}[ht!]
\centering
\caption{Absorption Lines Used for Sulfur and Iron Abundances}
\begin{tabular}{llllll}
\toprule
Ion & 12 + log(X/H)$_{\mathrm{LMC, tot}}$ & 12 + log(X/H)$_{\mathrm{SMC, tot}}$ & Wavelength (Å) & log $\lambda f_\lambda$ (Å) & Filter/Cenwave \\
\midrule
\multirow{2}{*}{S\,\textsc{ii}} & \multirow{2}{*}{6.94 $\pm$ 0.04 (1)} & \multirow{2}{*}{6.47 $\pm$ 0.03 (2)} & 1250.578 (3) & 0.809 (4) & G130M/1291, E140M/1425 \\
& & & 1253.805 (3) & 1.113 (4) & G130M/1291, E140M/1425 \\
\midrule
\multirow{6}{*}{Fe\,\textsc{ii}} & \multirow{6}{*}{7.32 $\pm$ 0.08 (5)(1)} & \multirow{6}{*}{6.89 $\pm$ 0.08 (5)} & 1142.366 (3) & 0.661 (3) & G130M/1291 \\
& & & 1143.226 (3) & 1.342 (3) & G130M/1291 \\
& & & 1144.938 (3) & 1.978 (3) & G130M/1291 \\
& & & 1608.451 (3) & 1.968 (3) & E140M/1425 \\
& & & 1611.201 (3) & 0.347 (3) & E140M/1425 \\
& & & 2249.877 (3) & 0.612 (3) & E230M/1978 \\
& & & 2260.780 (3) & 0.742 (3) & E230M/1978 \\
\bottomrule
\end{tabular}
\smallskip
\label{tab:lines-s-fe}

\textbf{Sources:} (1) Roman-Duval et al. (2021); (2) Jenkins \& Wallerstein (2017); (3) Morton (2003); (4) Kisielius et al. (2014); (5) Tchernyshyov et al. (2015)
\end{table*}

\begin{figure*}[ht!]
    
    \centering
    \includegraphics[width = 3.5in]{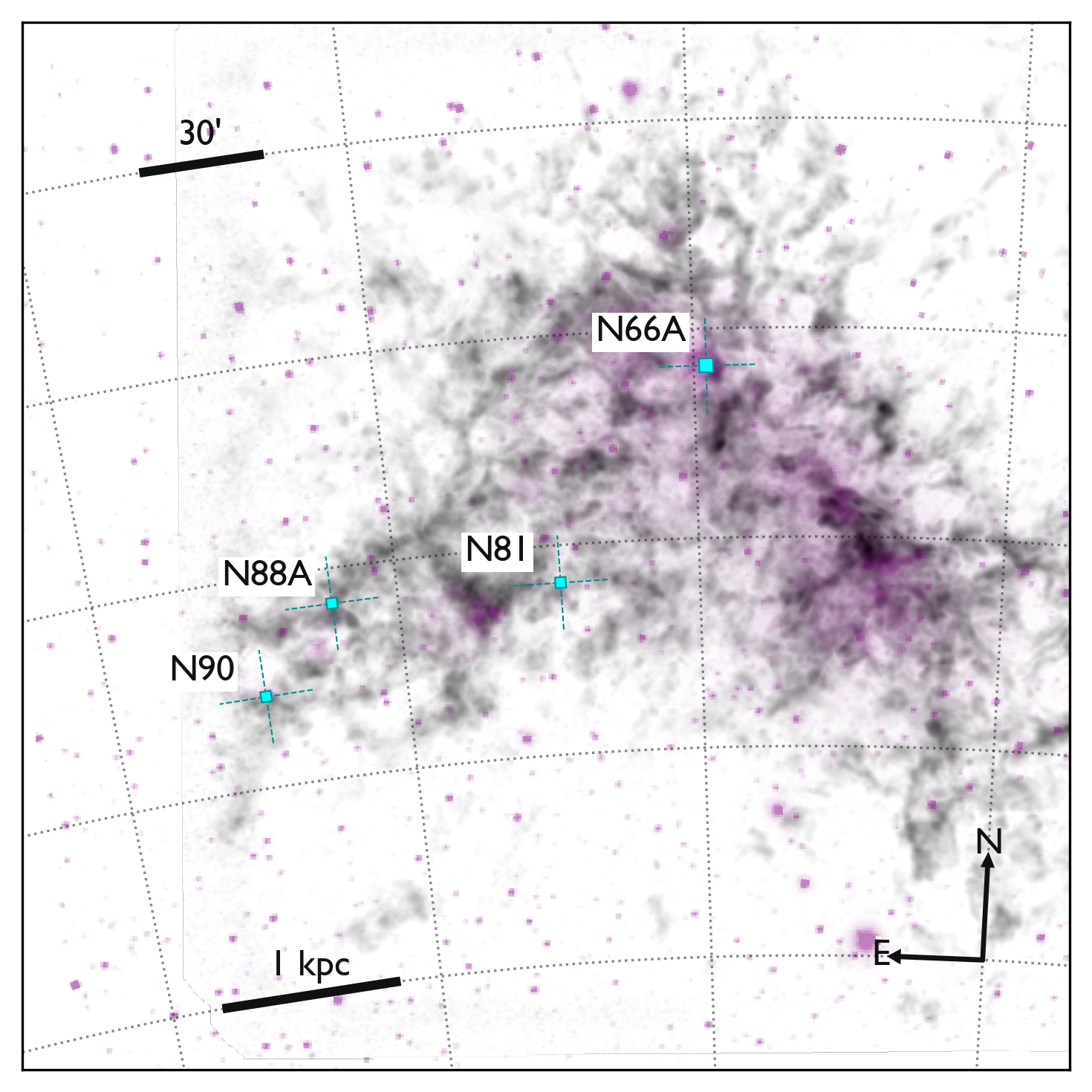}
    \includegraphics[width = 3.5in]{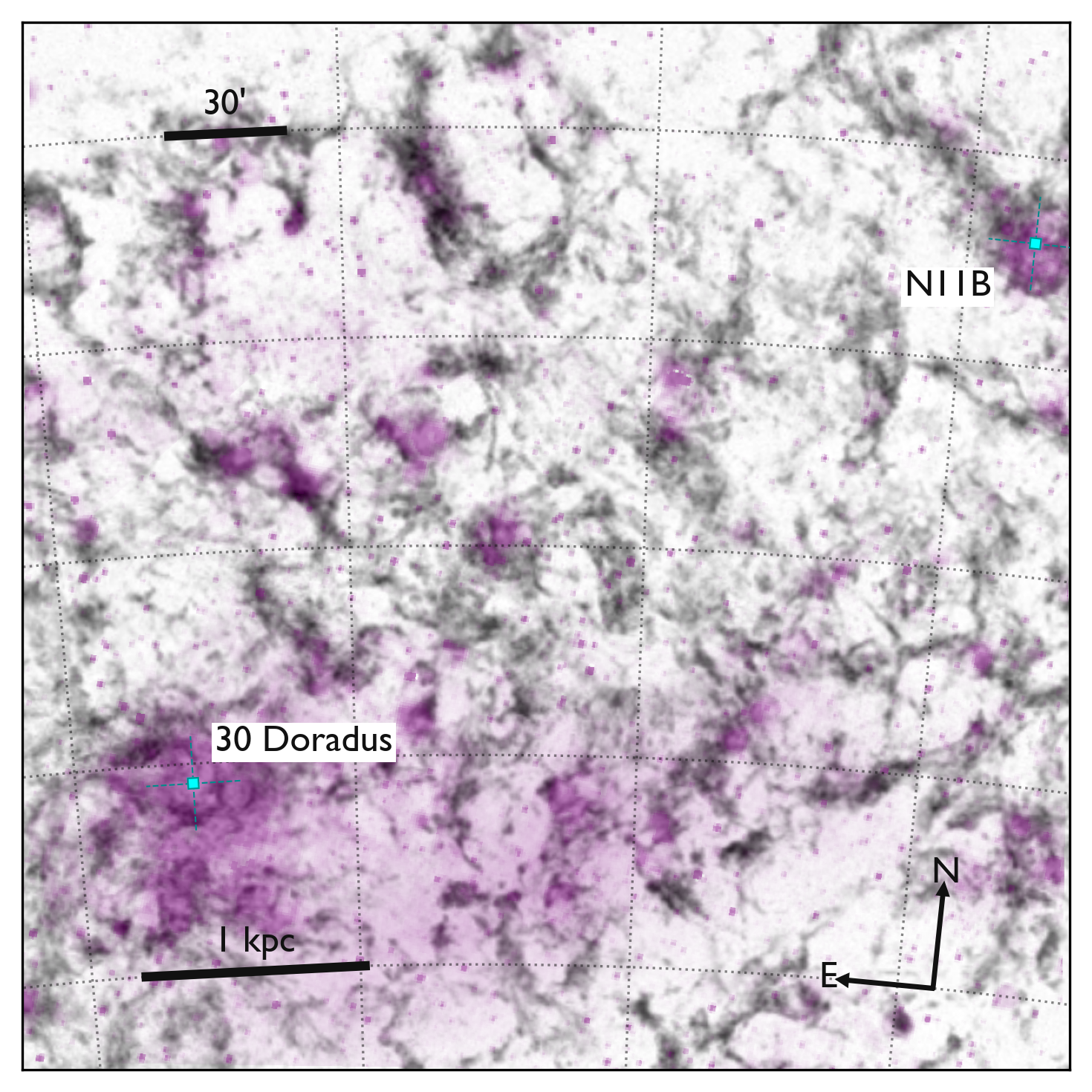}
    \caption{The location of H\,\textsc{ii} regions and sightlines included in this study relative to the Small Magellanic Cloud (SMC) and the Large Magellanic Cloud (LMC). The Australia Telescope Compact Array (ATCA) and Parkes HI survey peak 21 cm brightness temperature images \citep{stanimirovic_large-scale_1999, kim_neutral_2003} are shown in grayscale, while the Southern H$\alpha$ Sky Survey Atlas (SHASSA) \citep{gaustad_robotic_2001} images are shown in purple. Each cyan square contains both archival absorption spectra and literature emission spectra measurements, with zoomed-in fields of view presented in Figure~\ref{fig:smc-regions-sightlines}, Figure~\ref{fig:30-dor-sightlines}, and Figure~\ref{fig:n11b-sightlines}. The H\,\textsc{ii} regions 30 Doradus, LMC N11B, SMC N66A, SMC N81, SMC N88A, SMC N90 respectively correspond to 8, 9, 13, 1, 1, and 1 neutral gas sightlines. }
    \label{fig:smc-lmc-sightlines}
\end{figure*}

We have carefully selected sightlines that lie adjacent on the plane of the sky to emission-line observations by \citet{peimbert_chemical_2003, toribio_san_cipriano_carbon_2017, dominguez-guzman_homogeneity_2022}. For each H\,\textsc{ii} region in the listed studies, we gathered spectra from the MAST archive with angular separations of less than 3$^\prime$ from the center position of slits in the emission line observations, corresponding to $\sim54$ pc for the SMC ($d=62$ kpc) and $\sim44$ pc for the LMC ($d=50$ kpc). We visually inspect each of the sightlines using the Aladin Sky Atlas \citep{bonnarel_aladin_2000} to further verify their angular proximity to the H\,\textsc{ii} region (or a location within a more extended H\textsc{ii} region). The gathered UV spectral sightlines consist of observations by the UV Legacy Library of Young Stars as Essential Standards (ULLYSES) program \citep{roman-duval_uv_2025}, the Metal Evolution, Transport, and Abundance in the Large Magellanic Cloud (METAL) program \citep{roman-duval_metal_2019}, and miscellaneous observing programs with ID listed in Table~\ref{tab:sightlines}. Targets were observed with the Space Telescope Imaging Spectrograph (STIS) \citep{kimble_-orbit_1997, woodgate_space_1998} or the Cosmic Origins Spectrograph (COS) \citep{green_cosmic_2011} on the Hubble Space Telescope (HST). S/N along the sightlines ranges from $\sim$10 to 50, with a median of 20.6 for both SMC and LMC samples. Most of the 18 archival COS observations were carried out with both G130M and the G160M gratings, providing an approximate resolution $R\approx16,000$ and wavelength coverage of $\sim$1140--1780 Å. Three sightlines (Cl* NGC 346 MPG 782, Cl* NGC 2070 MEL 25, PGMW 3053) were observed with only the G130M grating with a central wavelength of 1291 Å (Cl* NGC 346 MPG 782 additionally observed with a central wavelength of 1096 Å) and a wavelength coverage of $\sim$1140--1420 Å. All STIS spectra were observed using the E140M grating ($R\approx45,800$, wavelength coverage $\sim$1200--1700 Å), with 3 of 17 sightlines (HD 5980, BI 42, PGMW 3120) having E230M spectra ($R\approx30,000$, wavelength coverage $\sim$1700--2300 Å) available that provide helpful, additional coverage of the NUV lines. One target HD 5980 is additionally observed using the E230M grating with a central wavelength of 2707 ($R\approx30,000$, wavelength coverage $\sim$2300--3100 Å) covering addition Fe\,\textsc{ii} absorption lines. Two sightlines are observed with both the G130M grating of COS and the E140M grating of STIS. Table~\ref{tab:sightlines} presents all spectra retrieved for this study. Figure~\ref{fig:smc-lmc-sightlines} depicts the location of the sightlines and the co-located H\,\textsc{ii} regions. 

Elemental abundances in several sightlines have been previously measured by \citet{jenkins_interstellar_2017} and \citet{roman-duval_metal_2021}. To reduce heterogeneity, we first perform measurements for all sightlines, then compare our measured values to previous results. This comparison is presented in Table~\ref{tab:comparison} of Section~\ref{compare-with-previous}. 

\startlongtable
\begin{deluxetable*}{cccccccccc}
\tablecaption{A List of Sightlines Measured in This Study \label{tab:sightlines}} 
\tabletypesize{\small}
\tablewidth{0pt}
\tablehead{
ID & HII Region & Target & R.A. & Decl. &
Sep.\ & Instr./Grat./Cen.\,$\lambda$. & $T_{\exp}$ & S/N & PID \\
& & & (h m s) & (d m s) & (') & & ('') & &
}
\startdata
A1 & SMC N66A & Cl* NGC 346 ELS 22 & 00 59 18.618 & -72 11 09.89 & 0.35 & COS/G130M/1291 & 360.000 & 27.03 & 11625 \\
 &  &  &  &  &  & COS/G130M/1327 & 360.032 & 28.34 & 11625 \\
 &  &  &  &  &  & COS/G160M/1577 & 560.032 & 15.81 & 11625 \\
 &  &  &  &  &  & COS/G160M/1623 & 560.032 & 14.97 & 11625 \\
A2 & SMC N66A & Cl* NGC 346 ELS 50 & 00 58 55.221 & -72 09 06.69 & 2.42 & COS/G130M/1096 & 9965.762 & 20.82 & 16103 \\
 &  &  &  &  &  & COS/G130M/1291 & 335.008 & 22.14 & 11625 \\
 &  &  &  &  &  & COS/G130M/1327 & 334.016 & 22.95 & 11625 \\
 &  &  &  &  &  & COS/G160M/1623 & 600.192 & 13.09 & 11625 \\
A3 & SMC N66A & Cl* NGC 346 ELS 51 & 00 59 08.697 & -72 10 14.14 & 0.92 & COS/G130M/1096 & 9959.808 & 19.22 & 16103 \\
 &  &  &  &  &  & COS/G130M/1291 & 360.032 & 21.99 & 11625 \\
 &  &  &  &  &  & COS/G130M/1327 & 360.032 & 22.67 & 11625 \\
 &  &  &  &  &  & COS/G160M/1577 & 648.224 & 13.66 & 11625 \\
 &  &  &  &  &  & COS/G160M/1623 & 648.224 & 12.84 & 11625 \\
A4 & SMC N66A & Cl* NGC 346 ELS 7 & 00 58 57.396 & -72 10 33.66 & 1.38 & STIS/E140M/1425 & 5400.000 & 16.68 & 7437 \\
A5 & SMC N66A & Cl* NGC 346 MPG 782 & 00 59 30.385 & -72 09 09.63 & 2.25 & COS/G130M/1096 & 10611.616 & 27.87 & 16371 \\
 &  &  &  &  &  & COS/G130M/1291 & 959.456 & 36.76 & 16371 \\
A6 & SMC N66A & Cl* NGC 346 NMC 17 & 00 59 06.750 & -72 10 41.26 & 0.68 & STIS/E140M/1425 & 9360.000 & 12.15 & 7437 \\
A7 & SMC N66A & Cl* NGC 346 NMC 28 & 00 59 01.819 & -72 10 31.22 & 1.09 & STIS/E140M/1425 & 6000.00 & 15.64 & 7437 \\
 &  &  &  &  &  & COS/G130M/1096 & 4394.784 & 22.25 & 16013 \\
A8 & SMC N66A & Cl* NGC 346 SSN 25 & 00 59 12.322 & -72 11 07.91 & 0.17 & STIS/E140M/1425 & 2707.000 & 8.78 & 15837 \\
A9 & SMC N66A & Cl* NGC 346 SSN 7 & 00 59 04.496 & -72 10 24.74 & 0.98 & STIS/E140M/1425 & 2210.200 & 21.43 & 16805 \\
A10 & SMC N66A & Cl* NGC 346 W 3 & 00 59 00.759 & -72 10 28.17 & 1.19 & STIS/E140M/1425 & 2215.200 & 17.79 & 16098 \\
 &  &  &  &  &  & STIS/E140M/1425 & 2567.200 & 18.99 & 16098 \\
 &  &  &  &  &  & STIS/E140M/1425 & 2700.000 & 21.19 & 7437 \\
A11 & SMC N66A & Cl* NGC 346 W 4 & 00 59 00.053 & -72 10 37.96 & 1.17 & STIS/E140M/1425 & 2707.000 & 15.25 & 15837 \\
A12 & SMC N66A & HD 5980 & 00 59 26.584 & -72 09 53.93 & 1.48 & STIS/E140M/1425 & 2028.000 & 35.58 & 14476 \\
 &  & (AzV 229, SK 78) &  &  &  & STIS/E140M/1425 & 2070.000 & 33.62 & 13373 \\
 &  &  &  &  &  & STIS/E140M/1425 & 2548.000 & 41.03 & 11623 \\
 &  &  &  &  &  & STIS/E140M/1425 & 2608.000 & 39.30 & 13373 \\
 &  &  &  &  &  & STIS/E140M/1425 & 2627.000 & 41.39 & 11623 \\
 &  &  &  &  &  & STIS/E140M/1425 & 2776.000 & 46.63 & 9094 \\
 &  &  &  &  &  & STIS/E140M/1425 & 2837.000 & 42.51 & 14476 \\
 &  &  &  &  &  & STIS/E140M/1425 & 3236.000 & 61.82 & 7480 \\
 &  &  &  &  &  & STIS/E230M/1978 & 1609.000 & 37.02 & 9094 \\
 &  &  &  &  &  & STIS/E230M/2707 & 1300.000 & 46.91 & 9094 \\
A13 & SMC N66A & SK 80 & 00 59 31.975 & -72 10 46.11 & 1.38 & STIS/E140M/1425 & 1200.000 & 17.97 & 9434 \\
B1 & SMC N81 & AzV 446 & 01 09 25.427 & -73 09 29.91 & 2.31 & COS/G130M/1291 & 323.008 & 31.89 & 11625 \\
 &  &  &  &  &  & COS/G130M/1327 & 323.040 & 33.21 & 11625 \\
 &  &  &  &  &  & COS/G160M/1577 & 630.208 & 20.87 & 11625 \\
 &  &  &  &  &  & COS/G160M/1623 & 620.192 & 20.47 & 11625 \\
C1 & SMC N88A & 2dFS 3694 & 01 24 34.445 & -73 09 08.88 & 1.9 & COS/G130M/1291 & 1534.400 & 56.54 & 16101 \\
 &  &  &  &  &  & COS/G160M/1611 & 2336.032 & 27.33 & 16101 \\
D1 & SMC N90 & SK 183 & 01 29 24.548 & -73 33 16.35 & 1.01 & STIS/E140M/1425 & 2707.000 & 17.78 & 15837 \\
 &  &  &  &  &  & COS/G130M/1096 & 1736.064 & 18.30 & 16808 \\
 &  &  &  &  &  & COS/G130M/1096 & 1736.096 & 17.54 & 17295 \\
E1 & 30 Doradus & BAT99 113 & 05 38 43.092 & -69 05 46.90 & 1.15 & COS/G130M/1291 & 2062.400 & 44.34 & 16812 \\
 &  &  &  &  &  & COS/G160M/1611 & 2004.032 & 20.88 & 16812 \\
E2 & 30 Doradus & BAT99 114 & 05 38 43.200 & -69 06 14.60 & 1.45 & COS/G130M/1291 & 2072.416 & 38.40 & 16812 \\
 &  &  &  &  &  & COS/G160M/1611 & 2024.096 & 17.49 & 16812 \\
E3 & 30 Doradus & Brey 77 & 05 38 42.104 & -69 05 55.34 & 1.3 & STIS/E140M/1425 & 2192.200 & 11.86 & 16090 \\
 &  & (BAT99 105) &  &  &  & STIS/E140M/1425 & 2638.199 & 12.95 & 16090 \\
 &  &  &  &  &  & STIS/E140M/1425 & 2638.200 & 12.50 & 16090 \\
E4 & 30 Doradus & Cl* NGC 2070 MEL 25 & 05 38 41.550 & -69 05 19.51 & 1.09 & COS/G130M/1291 & 2054.400 & 44.41 & 16810 \\
E5 & 30 Doradus & Cl* NGC 2070 MEL 47 & 05 38 37.718 & -69 05 20.97 & 1.44 & STIS/E140M/1425 & 2728.000 & 6.15 & 15629 \\
E6 & 30 Doradus & Cl* NGC 2070 MEL 55 & 05 38 33.975 & -69 04 21.23 & 1.91 & COS/G130M/1291 & 4774.368 & 34.43 & 16815 \\
 &  &  &  &  &  & COS/G160M/1611 & 4410.784 & 16.94 & 16815 \\
E7 & 30 Doradus & Cl* NGC 2070 MH 57 & 05 38 40.216 & -69 05 59.92 & 1.48 & COS/G130M/1291 & 2040.384 & 56.47 & 16093 \\
 &  &  &  &  &  & COS/G160M/1611 & 2080.064 & 27.19 & 16093 \\
E8 & 30 Doradus & RMC 140 & 05 38 41.616 & -69 05 13.96 & 1.08 & STIS/E140M/1425 & 2670.000 & 14.74 & 16272 \\
F1 & LMC N11B & BI 42 & 04 57 00.885 & -66 24 25.21 & 1.42 & STIS/E140M/1425 & 2240.000 & 10.84 & 14675 \\
 &  & (PGMW 3223) &  &  &  & STIS/E140M/1425 & 2750.000 & 11.40 & 14675 \\
 &  &  &  &  &  & STIS/E140M/1425 & 2778.000 & 11.61 & 14675 \\
 &  &  &  &  &  & STIS/E230M/1978 & 2240.000 & 13.77 & 14675 \\
 &  &  &  &  &  & STIS/E230M/1978 & 2730.000 & 14.60 & 14675 \\
F2 & LMC N11B & BRRG 75 & 04 56 43.276 & -66 25 02.49 & 0.55 & STIS/E140M/1425 & 2192.200 & 15.06 & 16820 \\
 &  &  &  &  &  & STIS/E140M/1425 & 2548.200 & 16.26 & 16820 \\
F3 & LMC N11B & PGMW 3053 & 04 56 41.048 & -66 24 40.54 & 0.59 & COS/G130M/1291 & 475.232 & 36.83 & 16095 \\
F4 & LMC N11B & PGMW 3058 & 04 56 42.134 & -66 24 54.71 & 0.55 & COS/G130M/1291 & 2066.368 & 44.27 & 16319 \\
 &  &  &  &  &  & COS/G160M/1611 & 1996.096 & 18.81 & 16319 \\
F5 & LMC N11B & PGMW 3100 & 04 56 45.203 & -66 25 10.78 & 0.57 & COS/G130M/1291 & 1990.400 & 36.25 & 16094 \\
 &  &  &  &  &  & COS/G160M/1611 & 1968.096 & 17.37 & 16094 \\
F6 & LMC N11B & PGMW 3120 & 04 56 46.798 & -66 24 46.86 & 0.15 & STIS/E140M/1425 & 2291.000 & 7.70 & 8320 \\
 &  &  &  &  &  & STIS/E140M/1425 & 3152.000 & 9.23 & 8320 \\
 &  &  &  &  &  & STIS/E230M/1978 & 1570.000 & 8.59 & 14675 \\
 &  &  &  &  &  & STIS/E230M/1978 & 2240.000 & 10.06 & 14675 \\
F7 & LMC N11B & PGMW 3168 & 04 56 54.460 & -66 24 15.87 & 0.84 & COS/G130M/1291 & 1970.400 & 53.38 & 16818 \\
 &  &  &  &  &  & COS/G160M/1611 & 1964.032 & 22.78 & 16818 \\
F8 & LMC N11B & PGMW 3204 & 04 56 58.790 & -66 24 40.71 & 1.19 & STIS/E140M/1425 & 2176.200 & 10.16 & 16815 \\
 &  &  &  &  &  & STIS/E140M/1425 & 2548.198 & 10.72 & 16815 \\
 &  &  &  &  &  & STIS/E140M/1425 & 2548.199 & 10.81 & 16815 \\
F9 & LMC N11B & [L72] LH 10-3061 & 04 56 42.507 & -66 25 18.22 & 0.8 & COS/G130M/1291 & 2066.368 & 42.26 & 16811 \\
 &  & (PGMW 3061) &  &  &  & COS/G160M/1611 & 1996.096 & 20.85 & 16811 \\
\enddata
\tablecomments{We designate each sightline with an ID, where the letter represents the corresponding H\,\textsc{ii} region, and the number denotes a specific target star. The sightline coordinates are retrieved from the SIMBAD database in the J2000 epoch. The ``Target'' column list the SIMBAD primary identifier, while other well-known identifiers are listed under. We list the per resolution element S/N calculated from the closest continuum segment from the central wavelength of each observation. }
\end{deluxetable*}

\subsubsection{Column Densities with Voigt Profile Fitting} \label{voigt}

\begin{figure*}[ht!]
    \centering
    \includegraphics[width = 6in]{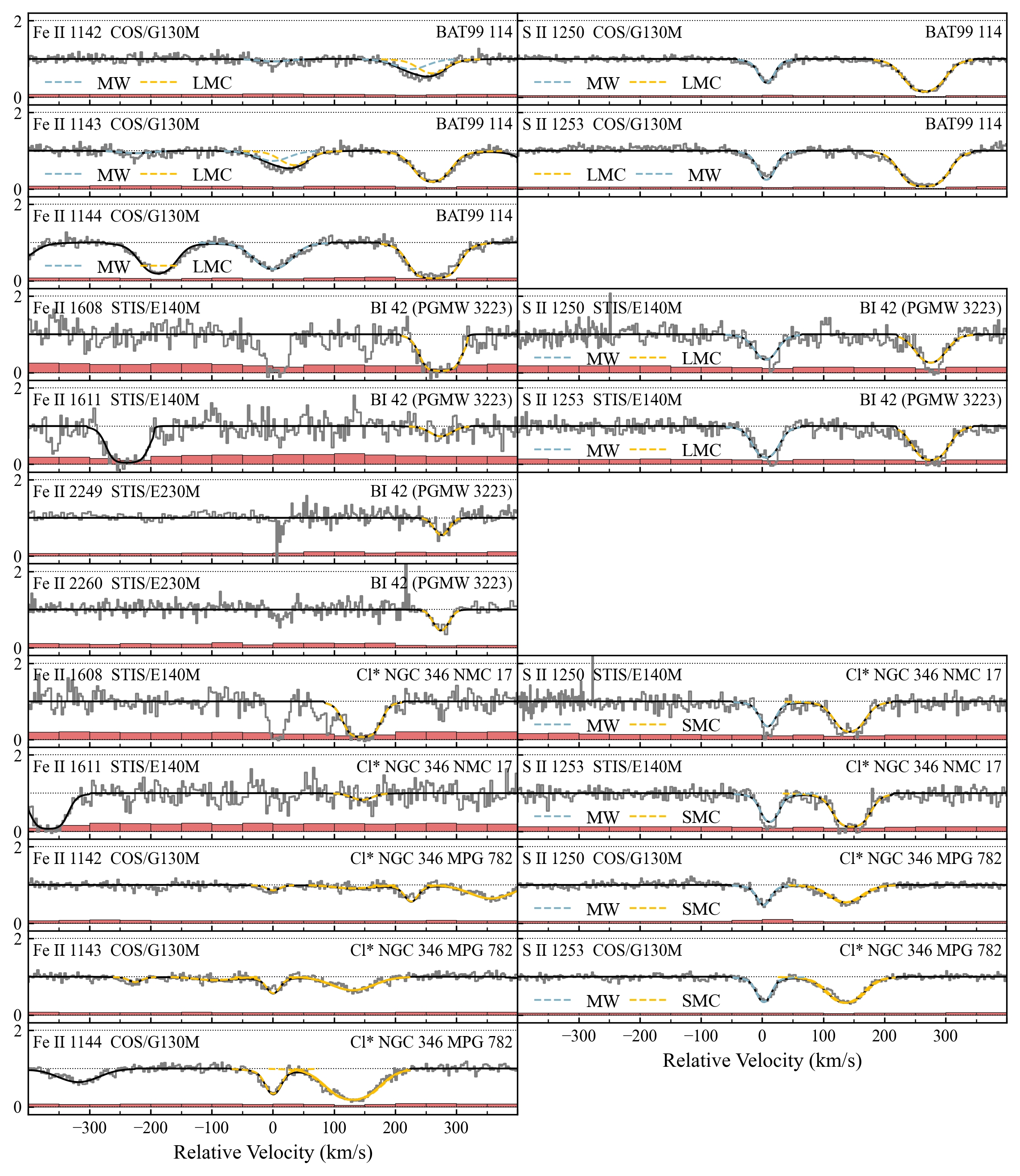}
    \caption{Examples of Voigt profile fitting for the Fe and S lines in this work. The Fe\,\textsc{ii} $\lambda\lambda\lambda$ 1142, 1143, 1144 Å and the S\,\textsc{ii} $\lambda\lambda$ 1250, 1253 Å lines are measured in the COS spectra for LMC target BAT 99 114 (corresponding to 30 Doradus) and SMC target 2dFS 3694 (corresponding to SMC N88A). The Fe\,\textsc{ii} ($\lambda\lambda$ 1608, 1611, 2249, 2260 Å) and the S\,\textsc{ii} $\lambda\lambda$ 1250, 1253 Å lines are measured in the STIS spectra for LMC target BI 42 (PGMW 3223, corresponding to LMC N11B) and SMC traget Cl* NGC 346 NMC 17 (corresponding to SMC N66A). Only the first spectra of each grating for each target is displayed here. The Milky Way components of the lines are shown in dashed light blue and the SMC/LMC components of the lines are shown in dashed light orange. The median uncertainty for each relative velocity range is shown in the light red bars on the bottom axes. We report the S abundances measured from the saturated S\,\textsc{ii} $\lambda\lambda$ 1250, 1253 Å lines as lower limits. }
    \label{fig:voigt}
\end{figure*}

As discussed in Section~\ref{sightlines}, we measure the neutral gas-phase elemental abundances of S and Fe via Voigt profile fitting. Continuum estimation is first performed for all the spectra in our dataset, using a combination of stellar model fitting \citep{tchernyshyov_ultraviolet_2025} and manual fitting. For stellar model fitting, we fit the PoWR model atmospheres for SMC/LMC OB and Wolf-Rayet stars \citep{sander_galactic_2012, todt_potsdam_2015, hainich_powr_2019} to wavelength ranges within our spectra free of ISM absorption lines. For cases where the stellar model fit is unsatisfactory, we proceed with the manual spline fitting tool \texttt{lt\_continuumfit} included in the spectral analysis package \texttt{Linetools} \citep{prochaska_linetoolslinetools_2016}. For fitting to the ISM absorption lines, we employ \texttt{the Veeper} \citep{burchett_veeper_2024} to simultaneously fit to all transitions of the same species, using guess parameters obtained manually using \texttt{pyigm IGMGuesses} \citep{prochaska_pyigmpyigm_2017}. The result for the S and Fe column densities are presented in Table~\ref{tab:abundances}. Example of the fitted Voigt profiles are shown in Figure~\ref{fig:voigt}. 

\subsubsection{Atomic and Molecular Hydrogen Column Density}

\begin{figure*}[ht!]
    \centering
    \includegraphics[width = 6in]{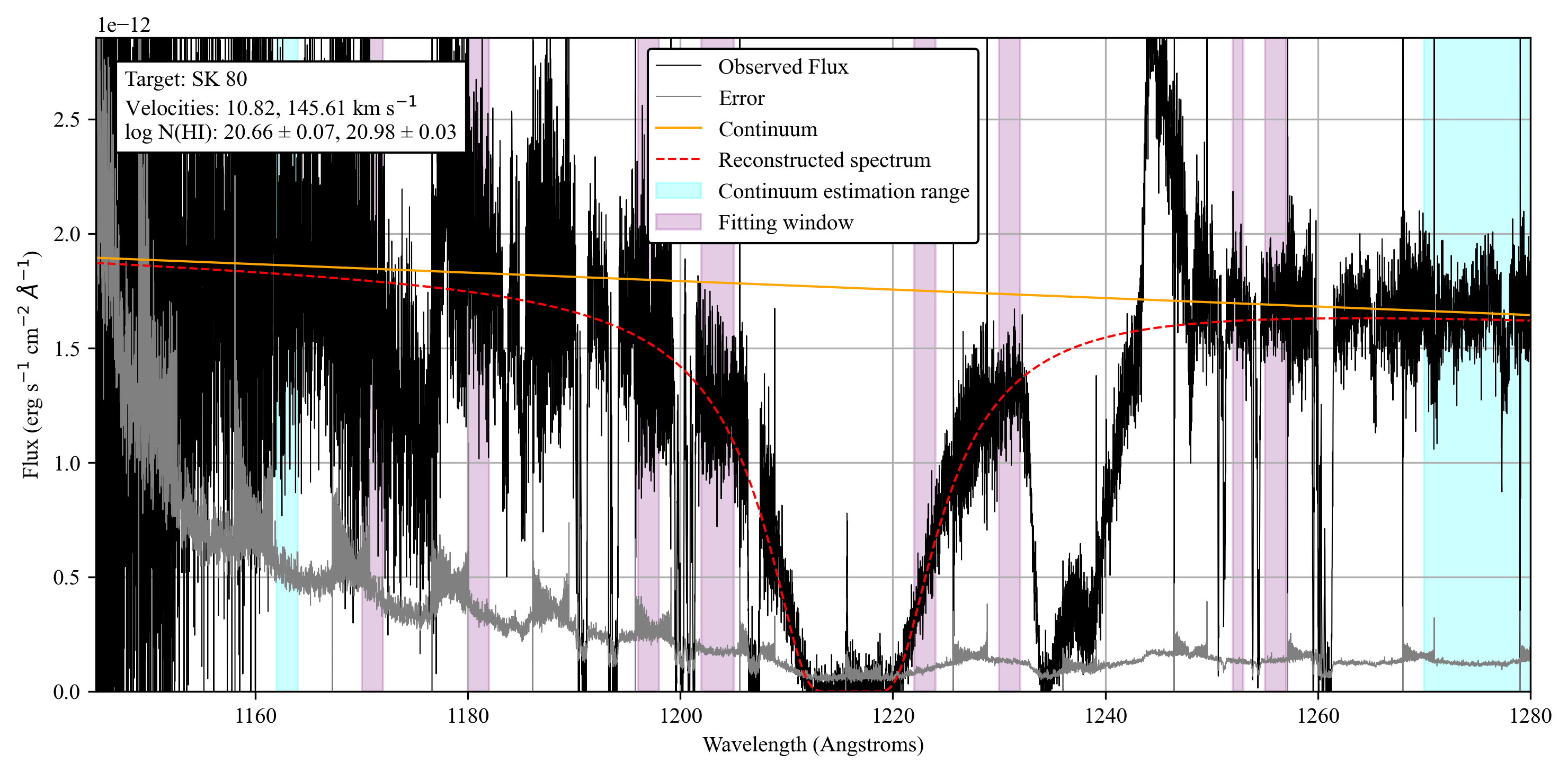}
    \caption{Fitting two convolved Lorentzian profiles to the Ly$\alpha$ absorption feature, using the SMC sightline SK 80 as an example. The linear continuum is shown in the solid yellow line. The parameters for the linear continuum fit are found by performing a least squares polynomial fit to the cyan regions of the spectra. The velocities to the Milky Way and the SMC/LMC components are estimated from fitting Voigt profiles to the S\,\textsc{ii} $\lambda\lambda$ 1250, 1253 Å absorption features. We sample the posterior distribution within the purple windows. The resulting atomic hydrogen column density is evaluated at the maximum posterior, where the 1$\sigma$ uncertainties are estimated from the 16th–84th percentile range of the posterior distribution. }
    \label{fig:nhi}
\end{figure*}

For the atomic hydrogen column density N(H\,\textsc{i}), we adopt the Ly$\alpha$ profile fitting method of \citet{diplas_iue_1994, roman-duval_metal_2019, hamanowicz_metal-z_2024}. As shown in Figure~\ref{fig:nhi}, we use a linear model to estimate the local continuum. We use the \texttt{polyfit} routine of the \texttt{numpy} \citep{van_der_walt_numpy_2011, harris_array_2020} \texttt{Python} package to find the best fitting first-order polynomial for two manually-selected ranges of wavelengths not covered by absorption or emission on either sides of the Ly$\alpha$ profile. The observed Ly$\alpha$ absorption feature consists of absorption from the Milky Way ISM and absorption from the Magellanic Clouds ISM. Using an approach similar to \citet{roman-duval_metal_2019}, we find the relative velocities of the Milky Way and the Magellanic Clouds components from jointly fitting the S\,\textsc{ii} $\lambda\lambda$ 1250, 1253 Å absorption profiles using \texttt{the Veeper} \citep{burchett_veeper_2024}. The continuum and the velocities are then used for forward modeling the Ly$\alpha$ absorption. 

We find the best fitting N(H\,\textsc{i})$_{\text{MW}}$ and N(H\,\textsc{i})$_{\text{MC}}$ using Markov Chain Monte Carlo (MCMC) sampling implemented in the \texttt{emcee} \citep{foreman-mackey_emcee_2013} \texttt{Python} package. For the Milky Way components, we adopt Gaussian priors in linear space that represent the range of values measured by \citet{welty_interstellar_2012} toward the SMC and the LMC. As \citet{welty_interstellar_2012} contain around one hundred sightlines for each galaxy, we adopt the Milky Way component column density range reported by \citet{welty_interstellar_2012} as the 3$\sigma$ range for the Gaussian prior. The ranges are listed in Table~\ref{tab:nhi-mw-prior}. 

\begin{table}[ht!]
\centering
\caption{Adopted Priors for the Milky Way Atomic Hydrogen Column Densities}
\begin{tabular}{lll}
\toprule
& Low 3$\sigma$ limit & High 3$\sigma$ limit \\
\midrule
SMC & $3.3\times10^{20}$ cm$^{-2}$ & $7.4\times10^{20}$ cm$^{-2}$ \\
LMC & $2.6\times10^{20}$ cm$^{-2}$ & $3.7\times10^{20}$ cm$^{-2}$ \\
\bottomrule
\end{tabular}
\smallskip
\label{tab:nhi-mw-prior}
\end{table}

We use uniform priors in linear space for log(NH\,\textsc{i})$_{\text{MC}}$ similar to \citet{roman-duval_metal_2019} (for the SMC between 18.0 and 22.0, for the LMC between 18.0 and 22.5). For each set of log(NH\,\textsc{i})$_{\text{MW}}$ and log(NH\,\textsc{i})$_{\text{MC}}$ and the pre-determined velocities, we construct the model spectrum by superimposing two Ly$\alpha$ profiles (given by Eq.~1 of \citet{roman-duval_metal_2019}) with the corresponding parameters, and convolving with the appropriate line-spread function (LSF). The log-likelihood is then evaluated for manually-selected ranges of the spectrum that are only absorbed by the Ly$\alpha$ transition. With \texttt{emcee}, we initialize 100 walkers from the priors and use 100 burn-in steps followed by 2000 production steps. The best-fit values for log(NH\,\textsc{i})$_{\text{MW}}$ and log(NH\,\textsc{i})$_{\text{MC}}$ are taken where the posterior is maximum, while the 1$\sigma$ uncertainties are estimated from the 16th–84th percentile range of the marginalized posterior. Uncertainties produced by this method do not account for errors introduced by the continuum fitting procedure, the selection of $\chi^2$ minimization ranges, or other assumptions employed to model the absorption profile. The result for the atomic hydrogen column densities are presented in Table~\ref{tab:abundances}. 

For molecular hydrogen column densities, one generally expects N(H$_2$) to be less than $10^{18}$ cm$^{-2}$ for Magellanic Clouds sightlines, which constitutes approximately 1$\%$ of the total N(H) \citep{welty_interstellar_2012}. \citet{welty_interstellar_2012} measured N(H$_2$) within seven sightlines toward the H\,\textsc{ii} regions included in this study, five of which are part of our neutral gas sample. Brey 77 (BAT99 105, Mk 42) toward 30 Doradus is reported to exhibit weak N(H$_2$) absorption of less than 1$\%$. Although not included in this study, Sk-69 243 and Sk-69 246, located $\sim0.6''$ and $\sim4'$ from the 30 Doradus slit position of \citet{peimbert_chemical_2003} respectively, exhibit weak N(H$_2$) absorption and a +0.015 dex (2.6$\%$) contribution from N(H$_2$) \citep{welty_interstellar_2012}. BI 42 (PGMW 3223) and PGMW 3120 toward LMC N11B are reported to have N(H$_2$) contributions of +0.0013 dex (0.3$\%$) and +0.0007 dex (0.16$\%$) respectively. HD 5980 (A12, AzV 229, SK 78) and SK 80 toward SMC N66A are reported to have log(NH$_2$) of 15.66 and 15.26 dex respectively, contributing negligible amounts ($<$0.0001 dex) to the total N(H). 

For the remaining sightlines, direct measurements of the molecular hydrogen contribution are not possible since the H$_2$ lines have not been observed. We consider the sightlines without N(H$_2$) measurements on a separate basis for each H\,\textsc{ii} region. The 30 Doradus sightlines exhibit the highest N(H\,\textsc{i}) values ranging from $3.16\times10^{21}$ cm$^{-2}$ to $2.00\times10^{22}$ cm$^{-2}$, beyond the threshold column density of $\sim5\times10^{21}$ cm$^{-2}$ for the transition to diffuse molecular ISM \citep{roman-duval_dust_2014, clark_quest_2023}. We notice the Cl\,\textsc{i} $\lambda$ 1347 and C\,\textsc{i} $\lambda\lambda$ 1158, 1656 Å absorption features are visible for all of the 30 Doradus sightlines, indicating the presence of molecular hydrogen along these lines of sight. We estimate the abundances of Cl\,\textsc{i} and C\,\textsc{i} toward all 30 Doradus sightlines using \texttt{linetools} \citep{prochaska_linetoolslinetools_2016} and find them comparable to those measured for Brey 77, which itself has measured weak N(H$_2$) absorption from \citet{welty_interstellar_2012}. Therefore, molecular hydrogen is very likely present and not negligible for the 30 Doradus sightlines. 

Cl\,\textsc{i} and C\,\textsc{i} absorption toward all LMC N11B sightlines and all SMC sightlines is weak. Additionally for these sightlines, the N(H$_2$) contribution is less of a concern given the lower average log(NH\,\textsc{i}) values of $\sim21.0$ dex. Based on Figure 17 of \citet{welty_interstellar_2012}, at such low levels of log(NH\,\textsc{i}), the sightlines will not contain substantial amounts of molecular hydrogen (if we assume N(H$_2$) is not significantly higher than N(H\,\textsc{i})). With the above considerations, we use a Gaussian distribution for log(NH$_2$) with a mean of 18.0 dex and a standard deviation of 2.0 dex when computing the total log(NH) for all sightlines. This effectively increases the uncertainty of the total log(NH) depending on the value we use for log(NH\,\textsc{i}). 

\subsection{\textsc{H}\,\textsc{ii} Region Elemental Abundances Adopted from Literature} \label{hii-abundances}

As discussed in Section~\ref{voigt}, we compare S and Fe abundances measured in the neutral gas with H\,\textsc{ii} region abundances from the literature. Abundances of S and Fe have been extensively carried out in H\,\textsc{ii} regions of the Magellanic Clouds. When selecting results from the literature, it is crucial to avoid introducing biases arising from differences in measurement techniques, uncertainty estimation methodologies, or assumed oscillator strengths. We select the recent, high-quality ground-based optical spectroscopy presented in \citet{toribio_san_cipriano_carbon_2017} and further analyzed by \citet{dominguez-guzman_homogeneity_2022} for S and Fe abundances in five H\,\textsc{ii} regions in the Magellanic Clouds. We include the observations of LMC H\,\textsc{ii} region 30 Doradus performed by \citet{peimbert_chemical_2003}, as they are presented in \citet{toribio_san_cipriano_carbon_2017} and \citet{dominguez-guzman_homogeneity_2022} along their own data as parallels. Fe abundances of 30 Doradus are adopted from \citet{rodriguez_fe_2005}'s analyses of \citet{peimbert_chemical_2003}'s data. Our final sample consists of six H\,\textsc{ii} regions with S and Fe abundance measurements summarized in Table~\ref{tab:hii-abundancess}.

\begin{table*}[ht!]
\centering
\caption{Summary of H\,\textsc{ii} Sulfur and Iron Abundances}
\begin{tabular}{lllll}
\toprule
Galaxy & H\,\textsc{ii} Region & 12 + log(S/H) & Adopted 12 + log(Fe/H) & Reference 12 + log(Fe/H) \\
\midrule
LMC & 30 Doradus & $6.99 \pm 0.10$ (1) & \colorbox{green!20}{$\leq 5.70$} (2) & $6.39 \pm 0.20$ (3); $5.90^{+0.06}_{-0.07}$ (4) \\
 & N11B & $6.72 \pm 0.02$ (5) & $5.43 \pm 0.02$ (6) & $5.43 \pm 0.02$ (7) \\
SMC & N66A & $6.35 \pm 0.03$ (5) & $5.26 \pm 0.02$ (6) & $5.17 \pm 0.03$ (7) \\
 & N81 & $6.34 \pm 0.03$ (5) & $5.48 \pm 0.02$ (6) & $5.26 \pm 0.03$ (7) \\
 & N88A & $6.33^{+0.10}_{-0.07}$ (5) & \colorbox{green!20}{$5.59 \pm 0.03$} (8) & $5.99 \pm 0.03$ (6); $5.52 \pm 0.02$ (7) \\
 & N90 & $6.38 \pm 0.04$ (5) & $<4.68$ (6) & $<4.72$ (7) \\
\bottomrule
\end{tabular}
\smallskip
\label{tab:hii-abundancess}

\textbf{Reference Fe abundances are listed for each H\,\textsc{ii} region to show method-dependent variations. Sources:} (1) \citet{peimbert_chemical_2003}; (2) \citet{rodriguez_fe_2005}, Fe\,\textsc{iii} + Fe\,\textsc{iv}, measured from \citet{peimbert_chemical_2003} spectra\footnote{reported as $\leq5.71$ in \citet{rodriguez_fe_2003}}; (3) \citet{peimbert_chemical_2003}, ionization corrected using \citet{relano_photoionization_2002} scheme; (4) \citet{rodriguez_fe_2005}, Fe\,\textsc{iii} measured from \citet{peimbert_chemical_2003} spectra, ionization corrected using \citet{rodriguez_fe_2005} Eq.~2; (5) \citet{dominguez-guzman_homogeneity_2022}, measured from \citet{toribio_san_cipriano_carbon_2017} spectra; (6) \citet{dominguez-guzman_homogeneity_2022}, Fe\,\textsc{iii} measured from \citet{toribio_san_cipriano_carbon_2017} spectra, ionization corrected using \citet{rodriguez_fe_2005} Eq.~2; (7) \citet{dominguez-guzman_homogeneity_2022}, Fe\,\textsc{iii} measured from \citet{toribio_san_cipriano_carbon_2017} spectra, ionization corrected using \citet{rodriguez_fe_2005} Eq.~3 and Eq.~4; (8) \citet{dominguez-guzman_homogeneity_2022}, Fe\,\textsc{iii} + Fe\,\textsc{iv}, measured from \citet{toribio_san_cipriano_carbon_2017} spectra.
\end{table*}

Ionization correction factors (ICFs) are required due to unobserved ions in H\,\textsc{ii} regions. The ICF is defined as the ratio between the total abundance of an element and the sum of its observed ions. In the case of Fe, the total abundance primarily consists of Fe\,\textsc{iii} (Fe$^{2+}$) and Fe\,\textsc{iv} (Fe$^{3+}$) \citep{rodriguez_fe_2005, mendez-delgado_gas-phase_2024}. While the Fe\,\textsc{iii} (Fe$^{2+}$) abundances are constrained by multiple emission lines \citep{peimbert_chemical_2003, toribio_san_cipriano_carbon_2017, dominguez-guzman_homogeneity_2022}, the Fe\,\textsc{iv} abundances are only constrained by the weak Fe\,\textsc{iv} $\lambda$6740 Å emission line. When Fe\,\textsc{iv} is unobserved, an ICF \citep[e.g.][]{rodriguez_fe_2005, izotov_chemical_2006} is required to calculate its abundance. For four of the H\,\textsc{ii} regions where Fe\,\textsc{iv} is unobserved (LMC N11B, SMC N66A, SMC N81, SMC N90), the total Fe abundances are calculated using Eq.~2 of \citet{rodriguez_fe_2005}, the photoionization modeling based ionization correction method preferred by recent studies \citep{dominguez-guzman_homogeneity_2022, mendez-delgado_gas-phase_2024}. Abundances calculated from other methods are listed in Table \ref{tab:hii-abundancess} for reference. With all the listed methods, the total Fe abundances are calculated from the observed Fe\,\textsc{iii} abundances and the ionization ratio of O\,\textsc{ii} to O\,\textsc{iii}. Fe\,\textsc{iv} is directly detected and measured in SMC N88A (therefore, negligible IC needed assuming low Fe\,\textsc{ii} and Fe\,\textsc{v}) and highlighted in green. Additionally, Fe\,\textsc{iv} is measured in 30 Doradus by \citet{rodriguez_fe_2005}, providing a total Fe abundance upper limit highlighted in green. The choices of Fe ICFs and related uncertainties are further discussed in Section~\ref{ic}. For S, given both S\,\textsc{ii} and S\,\textsc{iii} are detected for all the H\,\textsc{ii} regions in the sample, the ICF only needs to address marginal contributions from S\,\textsc{iv} and higher ionized S species ($\sim\pm0.1$ dex \citep{amayo_ionization_2021}). For the S abundances reported by \citet{peimbert_chemical_2003}, ICF from \citet{garnett_abundance_1989} was employed; for all other reported S abundances, ICF from \citet{amayo_ionization_2021} was employed. 

\section{Results} \label{results}

\subsection{Comparing the \textsc{H}\,\textsc{ii} Region and Neutral Gas Abundances} \label{hi-vs-hii}

Table~\ref{tab:abundances} reports the measured atomic hydrogen, sulfur (S), and iron (Fe) column densities, using methods described in Section~\ref{methods}. 

\begin{deluxetable*}{ccccccc}
\tablecaption{Measured S and Fe Absorption-Line Column Densities and Abundances \label{tab:abundances}}
\tabletypesize{\small}
\tablewidth{0pt}
\tablehead{
\colhead{ID} & \colhead{Target} & \colhead{$\log(\mathrm{N\,H\textsc{i}})_\text{MC}$} & \colhead{$\log(\mathrm{N\,S\textsc{ii}})_\text{MC}$} & \colhead{$\log(\mathrm{N\,Fe\textsc{ii}})_\text{MC}$} & \colhead{$12 + \log(\mathrm{S/H})_\text{MC}$} & \colhead{$12 + \log(\mathrm{Fe/H})_\text{MC}$}
}
\startdata
A1 & Cl* NGC 346 ELS 22 & $21.176 \pm 0.017$ & $15.567 \pm 0.013$ & $15.114 \pm 0.138$ & $6.390 \pm 0.022$ & $5.937 \pm 0.139$ \\
A2 & Cl* NGC 346 ELS 50 & $20.816 \pm 0.040$ & $15.403 \pm 0.011$ & $14.662 \pm 0.008$ & $6.585 \pm 0.042$ & $5.844 \pm 0.041$ \\
A3 & Cl* NGC 346 ELS 51 & $20.905 \pm 0.033$ & $15.482 \pm 0.010$ & $14.853 \pm 0.275$ & $6.576 \pm 0.035$ & $5.947 \pm 0.277$ \\
A4 & Cl* NGC 346 ELS 7 & $21.096 \pm 0.022$ & $15.602 \pm 0.018$ & $15.218 \pm 0.081$ & $6.505 \pm 0.029$ & $6.121 \pm 0.084$ \\
A5 & Cl* NGC 346 MPG 782 & $20.043 \pm 0.157$ & $15.366 \pm 0.008$ & $14.767 \pm 0.041$ & $7.315 \pm 0.158$ & $6.716 \pm 0.163$ \\
A6 & Cl* NGC 346 NMC 17 & $21.288 \pm 0.016$ & $15.791 \pm 0.035$ & $15.170 \pm 0.116$ & $6.502 \pm 0.039$ & $5.881 \pm 0.117$ \\
A7 & Cl* NGC 346 NMC 28 & $21.030 \pm 0.026$ & $15.567 \pm 0.021$ & $14.737 \pm 0.007$ & $6.536 \pm 0.034$ & $5.706 \pm 0.027$ \\
A8 & Cl* NGC 346 SSN 25 & $21.487 \pm 0.019$ & $15.686 \pm 0.027$ & $14.976 \pm 0.099$ & $6.199 \pm 0.033$ & $5.489 \pm 0.101$ \\
A9 & Cl* NGC 346 SSN 7 & $20.962 \pm 0.024$ & $15.511 \pm 0.010$ & $14.876 \pm 0.023$ & $6.548 \pm 0.027$ & $5.913 \pm 0.034$ \\
A10 & Cl* NGC 346 W 3 & $21.048 \pm 0.013$ & $15.512 \pm 0.008$ & $14.952 \pm 0.059$ & $6.463 \pm 0.016$ & $5.903 \pm 0.061$ \\
A11 & Cl* NGC 346 W 4 & $21.019 \pm 0.026$ & $15.624 \pm 0.020$ & $14.844 \pm 0.027$ & $6.604 \pm 0.033$ & $5.824 \pm 0.038$ \\
A12 & HD 5980 & $21.057 \pm 0.031$ & $15.478 \pm 0.004$ & $14.918 \pm 0.026$ & $6.704 \pm 0.035$ & $6.144 \pm 0.043$ \\
A13 & SK 80 & $20.983 \pm 0.030$ & $15.529 \pm 0.012$ & $15.188 \pm 0.065$ & $6.545 \pm 0.032$ & $6.204 \pm 0.072$ \\
B1 & AzV 446 & $21.168 \pm 0.019$ & $15.613 \pm 0.010$ & $15.290 \pm 0.036$ & $6.444 \pm 0.022$ & $6.121 \pm 0.041$ \\
C1 & 2dFS 3694 & $20.779 \pm 0.020$ & $15.646 \pm 0.120$ & $14.392 \pm 0.011$ & $6.865 \pm 0.122$ & $5.611 \pm 0.023$ \\
D1 & SK 183 & $20.595 \pm 0.074$ & $15.353 \pm 0.013$ & $14.662 \pm 0.013$ & $6.756 \pm 0.075$ & $6.065 \pm 0.075$ \\
E1 & BAT99 113 & $21.785 \pm 0.003$ & $>15.851$ & $15.383 \pm 0.017$ & $>6.066$ & $5.598 \pm 0.017$ \\
E2 & BAT99 114 & $21.871 \pm 0.002$ & $>15.921$ & $15.381 \pm 0.016$ & $>6.050$ & $5.510 \pm 0.016$ \\
E3 & Brey 77 & $21.764 \pm 0.015$ & $>16.685$ & $15.737 \pm 0.025$ & $>6.921$ & $5.973 \pm 0.029$ \\
E4 & Cl* NGC 2070 MEL 25 & $21.788 \pm 0.003$ & $>15.944$ & $15.485 \pm 0.016$ & $>6.156$ & $5.697 \pm 0.016$ \\
E5 & Cl* NGC 2070 MEL 47 & $22.000 \pm 0.001$ & $>16.109$ & $15.479 \pm 0.135$ & $>6.109$ & $5.479 \pm 0.135$ \\
E6 & Cl* NGC 2070 MEL 55 & $21.702 \pm 0.004$ & $>15.998$ & $15.548 \pm 0.023$ & $>6.295$ & $5.845 \pm 0.023$ \\
E7 & Cl* NGC 2070 MH 57 & $21.885 \pm 0.002$ & $>16.127$ & $15.440 \pm 0.010$ & $>6.242$ & $5.555 \pm 0.010$ \\
E8 & RMC 140 & $21.936 \pm 0.007$ & $>16.292$ & $15.738 \pm 0.044$ & $>6.356$ & $5.802 \pm 0.045$ \\
F1 & BI 42 & $21.457 \pm 0.006$ & $>15.694$ & $15.245 \pm 0.037$ & $>6.237$ & $5.788 \pm 0.038$ \\
F2 & BRRG 75 & $21.520 \pm 0.007$ & $>15.994$ & $15.624 \pm 0.031$ & $>6.474$ & $6.104 \pm 0.032$ \\
F3 & PGMW 3053 & $21.323 \pm 0.004$ & $>15.605$ & $14.994 \pm 0.018$ & $>6.282$ & $5.671 \pm 0.019$ \\
F4 & PGMW 3058 & $21.598 \pm 0.003$ & $>15.622$ & $14.924 \pm 0.016$ & $>6.024$ & $5.326 \pm 0.016$ \\
F5 & PGMW 3100 & $21.687 \pm 0.003$ & $>15.599$ & $15.029 \pm 0.023$ & $>5.912$ & $5.342 \pm 0.023$ \\
F6 & PGMW 3120 & $21.498 \pm 0.016$ & $>15.828$ & $15.240 \pm 0.059$ & $>6.330$ & $5.742 \pm 0.061$ \\
F7 & PGMW 3168 & $21.472 \pm 0.003$ & $>15.682$ & $15.311 \pm 0.292$ & $>6.210$ & $5.839 \pm 0.292$ \\
F8 & PGMW 3204 & $21.419 \pm 0.032$ & $>15.695$ & $14.962 \pm 0.026$ & $>6.275$ & $5.542 \pm 0.041$ \\
F9 & [L72] LH 10-3061 & $21.561 \pm 0.004$ & $>15.746$ & $14.876 \pm 0.017$ & $>6.185$ & $5.315 \pm 0.017$ \\
\enddata
\end{deluxetable*}

Three of the H\,\textsc{ii} regions in our sample are associated with multiple neutral gas sightlines. For each sightline, the individual measurement errors resulting from uncertainty propagation are small ($\sim0.05$ dex), compared to the dispersion in 12 + log(X/H) between sightlines ($\sim0.2$ dex). This could be the intrinsic heterogeneity of the neutral ISM elemental abundance and depletion \citep[e.g.][]{jenkins_unified_2009, tchernyshyov_elemental_2015, jenkins_interstellar_2017, roman-duval_metal_2021}, or other factors such as unaccounted for continuum fitting uncertainties. We compute the mean abundance, standard error of the mean, and the intrinsic spread due to the above factors using the Bayesian parameter estimation scheme presented in Section 5.6 of \citet{ivezic_statistics_2014}. Table~\ref{tab:hi-abundances-summary} reports the mean abundance, standard error of the mean, and the intrinsic spread derived from all neutral gas measurements. 

\begin{table*}[ht!]
\centering
\caption{Summarized Sulfur and Iron Abundances for Neutral Gas Surrounding Each H\,\textsc{ii} Region}
\begin{tabular}{llll}
\toprule
Galaxy & H\,\textsc{ii} Region & Mean 12 + log(S/H) & Mean 12 + log(Fe/H) \\
\midrule
LMC & 30 Doradus & $>6.21 \pm 0.04 \pm 0.10$ & $5.69 \pm 0.05 \pm 0.14$ \\
 & N11B & $>6.21 \pm 0.05 \pm 0.14$ & $5.61 \pm 0.09 \pm 0.26$ \\
SMC & N66A & $6.54 \pm 0.03 \pm 0.12$ & $5.95 \pm 0.06 \pm 0.18$ \\
 & N81 & $6.44 \pm 0.02$ & $6.12 \pm 0.04$ (single sightline) \\
 & N88A & $6.87 \pm 0.12$ & $5.61 \pm 0.02$ (single sightline) \\
 & N90 & $6.76 \pm 0.08$ & $6.07 \pm 0.08$ (single sightline) \\
\bottomrule
\end{tabular}
\tablecomments{For H\,\textsc{ii} regions with multiple sightlines, the first uncertainty is the standard error of the mean 12 + log(X/H) abundance, and the second uncertainty is the intrinsic spread between the sightlines. }
\smallskip
\label{tab:hi-abundances-summary}
\end{table*}

\begin{figure*}[ht!]
    \centering
    \includegraphics[width = 6in]{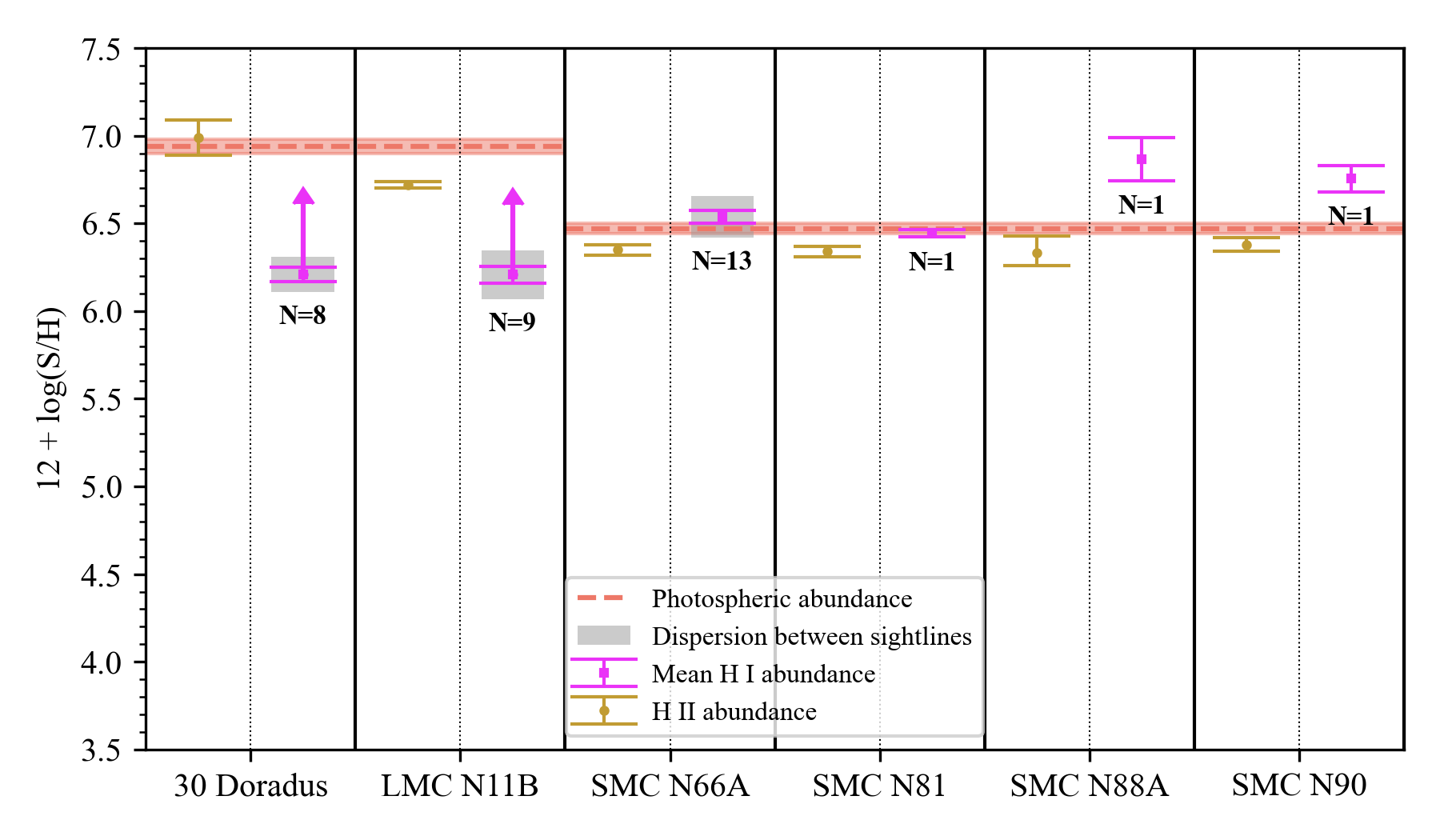}
    \caption{A side-by-side comparison of 12 + log(S/H) in neutral gas and adjacent H\,\textsc{ii} regions. The two left panels show the LMC H\,\textsc{ii} regions, while the four right panels show the SMC H\,\textsc{ii} regions. The left column of each panel shows the literature H\,\textsc{ii} S region abundance detailed in Section~\ref{hii-abundances}. The right columns of each panel show the measured neutral sightline S abundances/lower limits, with N denoting the number of corresponding UV absorption sightlines. The purple error bars denote only the standard error of the mean neutral gas S abundance, while the gray bands denote the sightline-to-sightline dispersion. In the right columns of the first two panels (representing LMC sightlines), we provide the lower limit of the S abundances in neutral gas due to the effects of unresolved saturation discussed in \citep{roman-duval_metal_2021} and can be seen in Figure~\ref{fig:voigt} for STIS observations. In the SMC, the overall metallicity is lower; absorption features are less affected by saturation \citep{jenkins_interstellar_2017}. Thus, neutral gas S abundances are not listed as limits.}
    \label{fig:s-abundances}
\end{figure*}

\begin{figure*}[ht!]
    \centering
    \includegraphics[width = 6in]{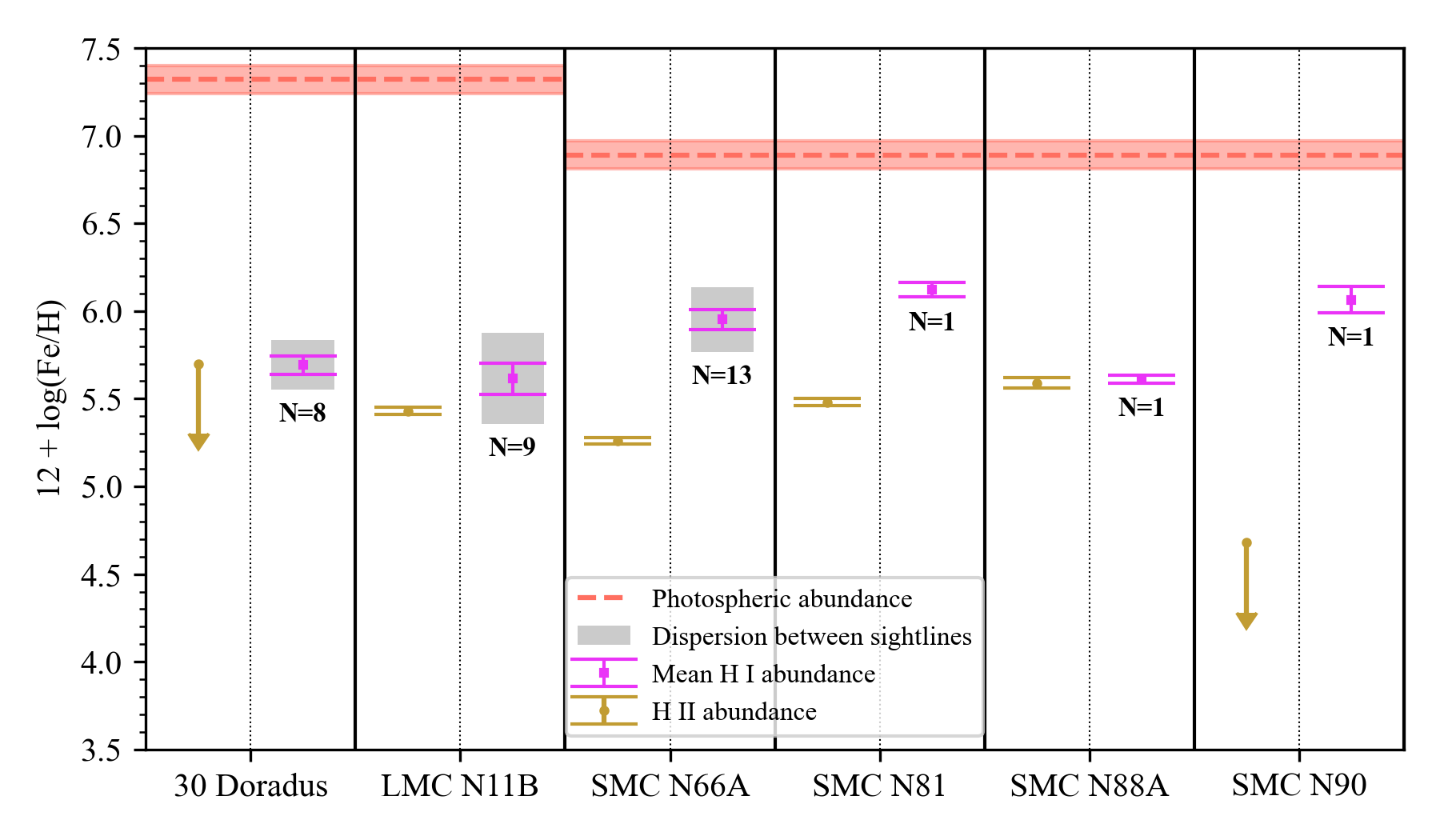}
    \caption{A side-by-side comparison of 12 + log(Fe/H) in neutral gas and adjacent H\,\textsc{ii} regions. The two left panels show the LMC H\,\textsc{ii} regions, while the four right panels show the SMC H\,\textsc{ii} regions. As in the previous Figure for S, the left column of each panel shows the adopted literature H\,\textsc{ii} region Fe abundances, listed in Table~\ref{tab:hii-abundancess} as the boxed 12 + log(Fe/H) values. The 30 Doradus Fe abundance upper limit and the SMC N88A Fe abundance are sums of measured Fe\,\textsc{iii} and Fe\,\textsc{iv}. The remaining abundances are inferred from ionization corrections using Eq.~2 of \citet{rodriguez_fe_2005}. Additional details of the employed H\,\textsc{ii} region Fe abundances are discussed in in Section~\ref{hii-abundances}. The right columns show the measured neutral sightline Fe abundances, with N denoting the number of sightlines used in calculation of the mean. In the right columns, the gray shaded regions represent the intra-sightline dispersion (when there are multiple sightlines), while the purple error bars represent the mean and its uncertainty. A large percentage of gas-phase Fe is depleted into dust grains in neutral gas (as expected from established studies \citep{jenkins_interstellar_2017, roman-duval_metal_2019, roman-duval_metal_2021, roman-duval_metal_2022, roman-duval_metal_2022-1}).}
    \label{fig:fe-abundances}
\end{figure*}

\begin{figure*}[ht!]
    \centering
    \includegraphics[width = 6in]{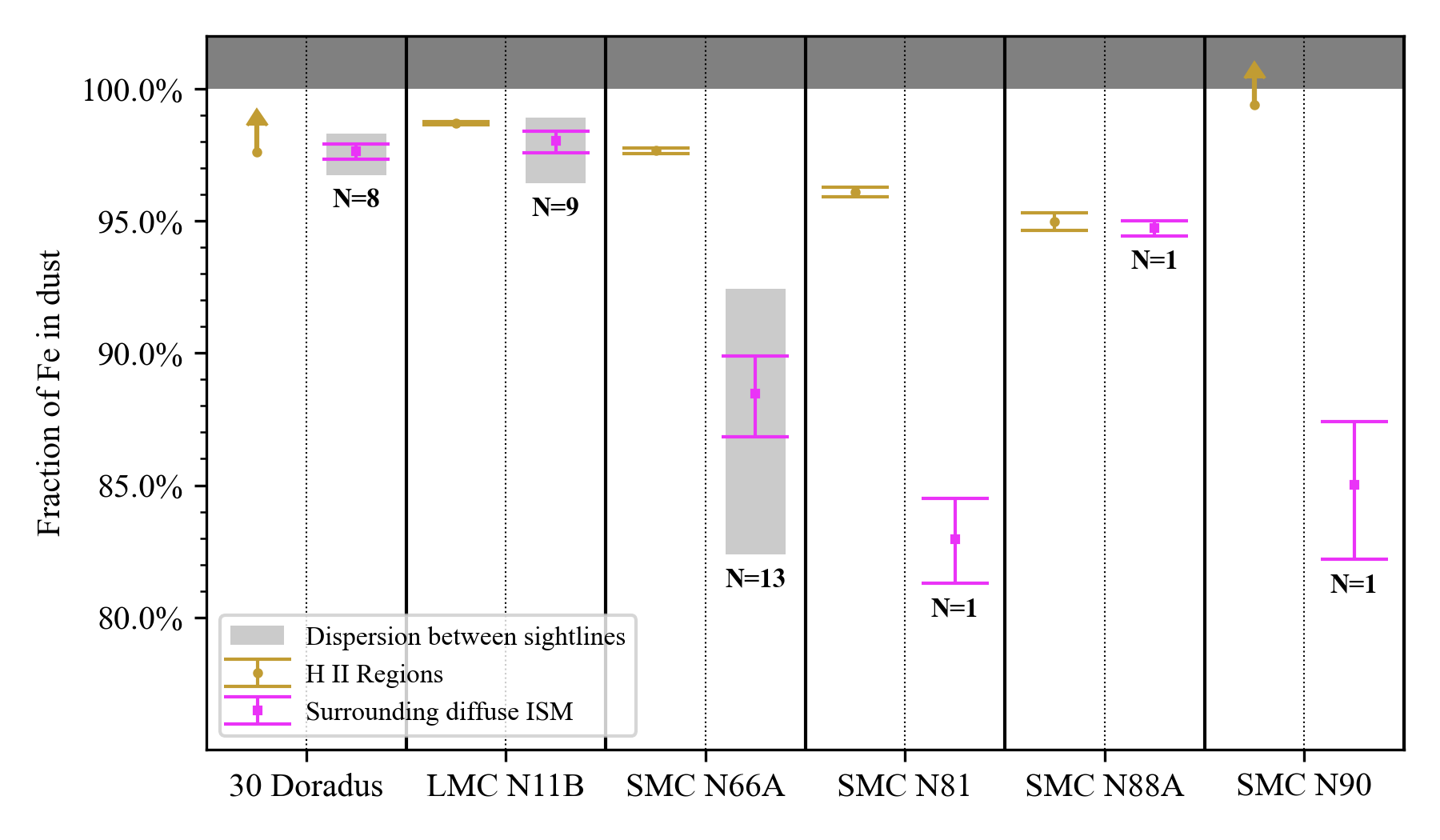}
    \caption{A side-by-side comparison of the fraction of Fe depleted into dust in neutral gas and adjacent H\,\textsc{ii} regions. The two left panels show the LMC H\,\textsc{ii} regions, while the four right panels show the SMC H\,\textsc{ii} regions. The left and right columns respectively show the percentage of Fe depleted into dust for the H\,\textsc{ii} region gas and the neighboring neutral ISM. In the right columns, N denotes the number of neutral gas sightlines. The gray bands represent the intra-sightline dispersion, while the purple error bars represent dust fraction derived from the mean depletion.}
    \label{fig:fe-in-dust}
\end{figure*}

Figure~\ref{fig:s-abundances} shows the side-by-side comparison of S abundances in the H\,\textsc{ii} regions and surrounding neutral gas. Figure~\ref{fig:fe-abundances} shows a similar plot for the Fe abundances in the H\,\textsc{ii} regions and surrounding neutral gas. Figure~\ref{fig:fe-in-dust} shows the fraction of Fe depleted into dust for the H\,\textsc{ii} regions and surrounding neutral gas. 

\begin{figure*}[ht!]
    \centering
    \includegraphics[width = 6in]{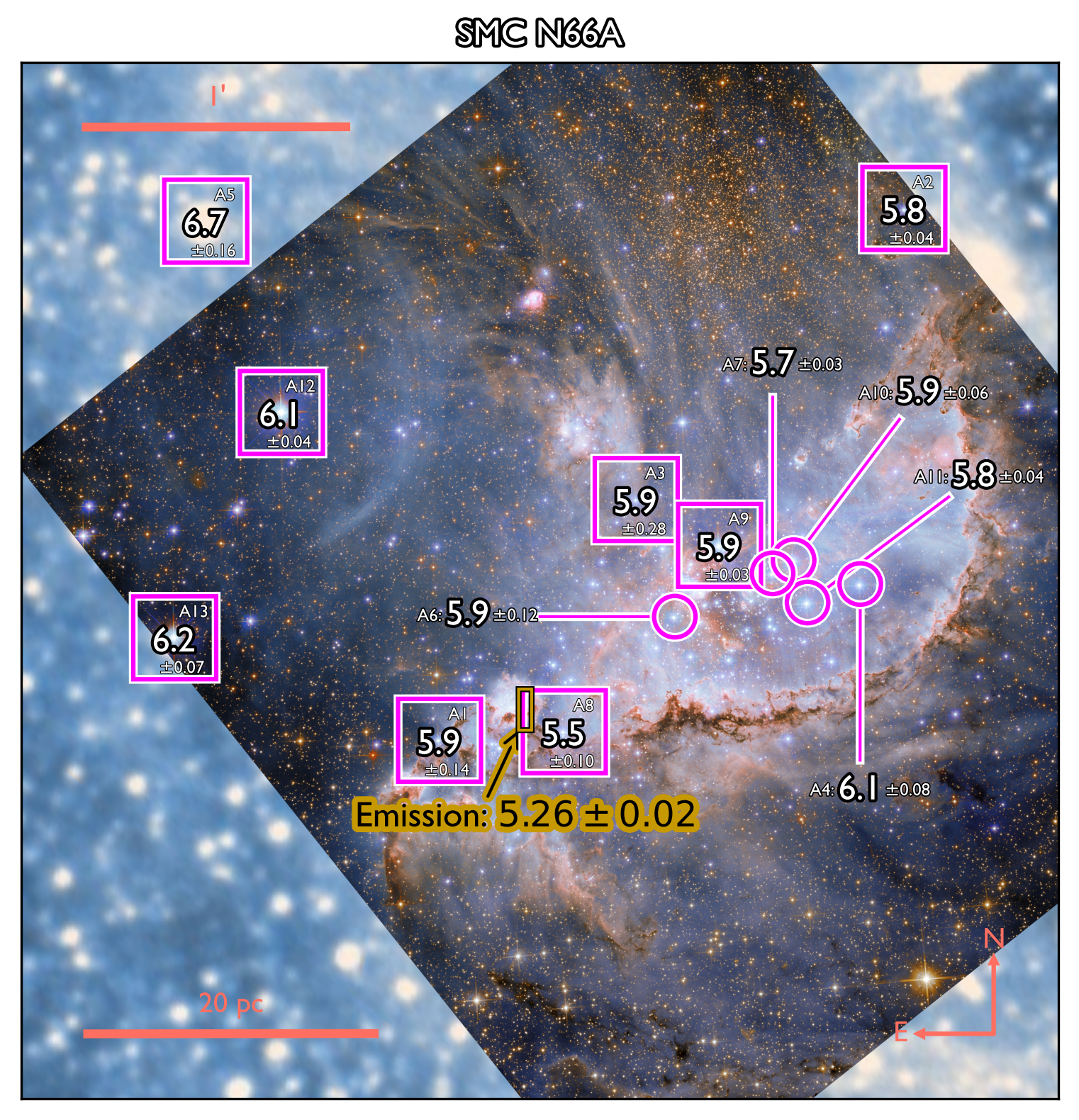}
    \includegraphics[width = 2in]{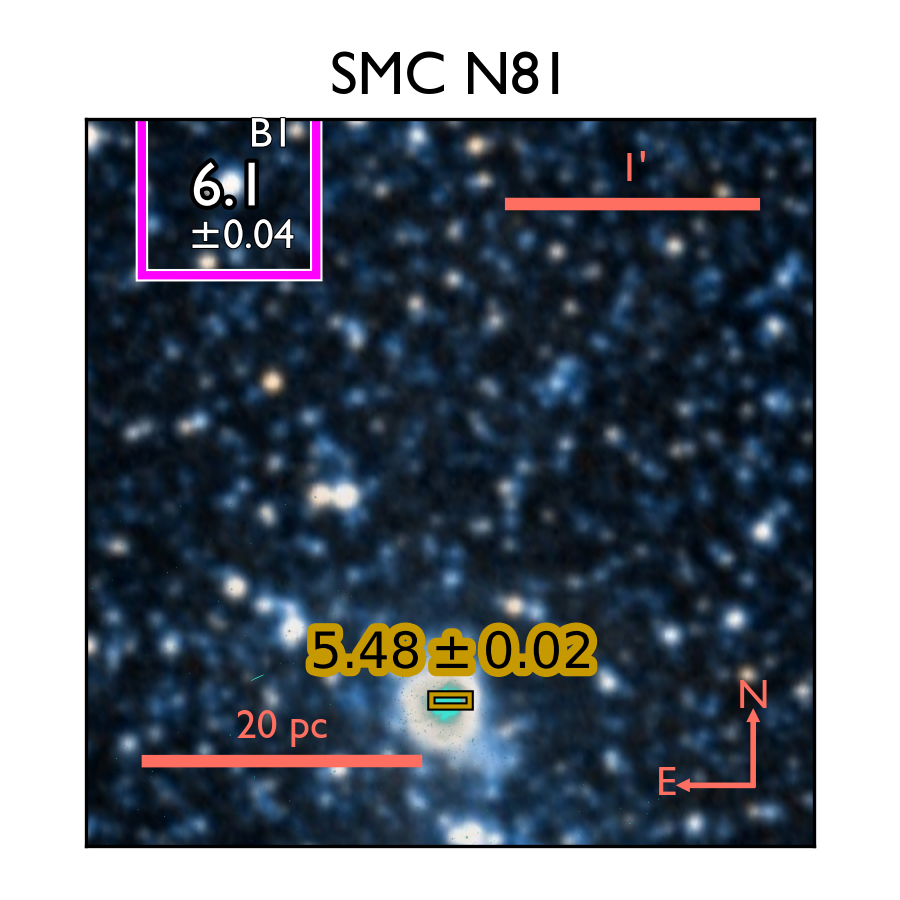}
    \includegraphics[width = 2in]{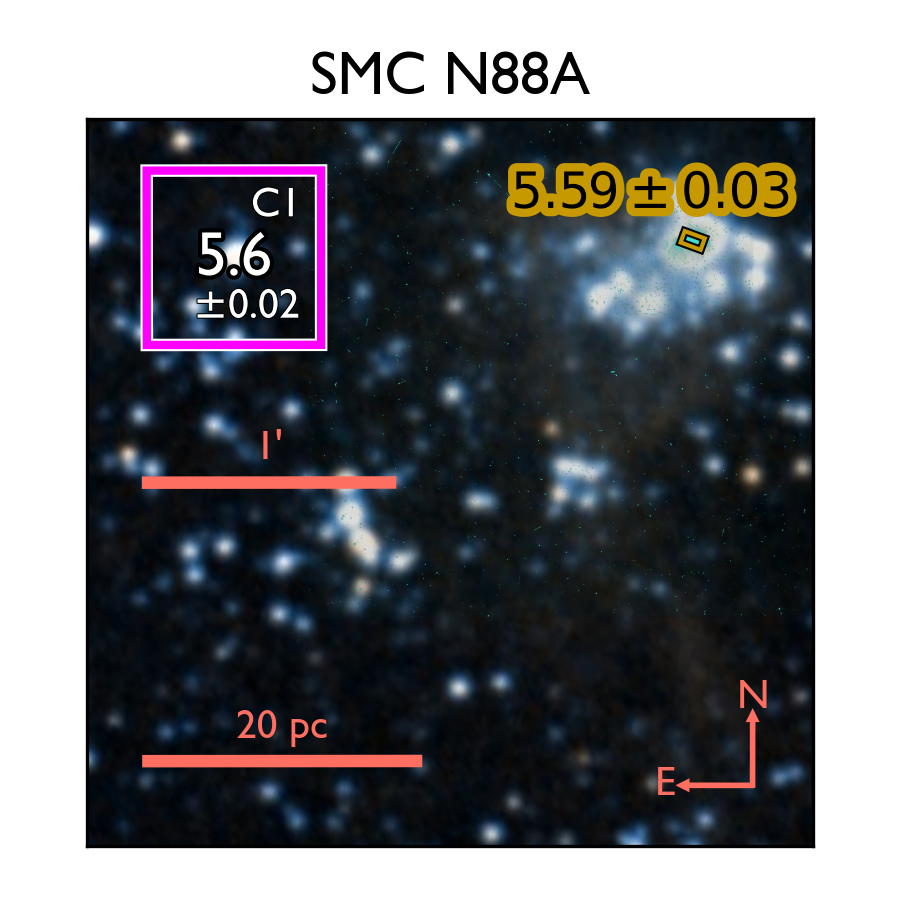}
    \includegraphics[width = 2in]{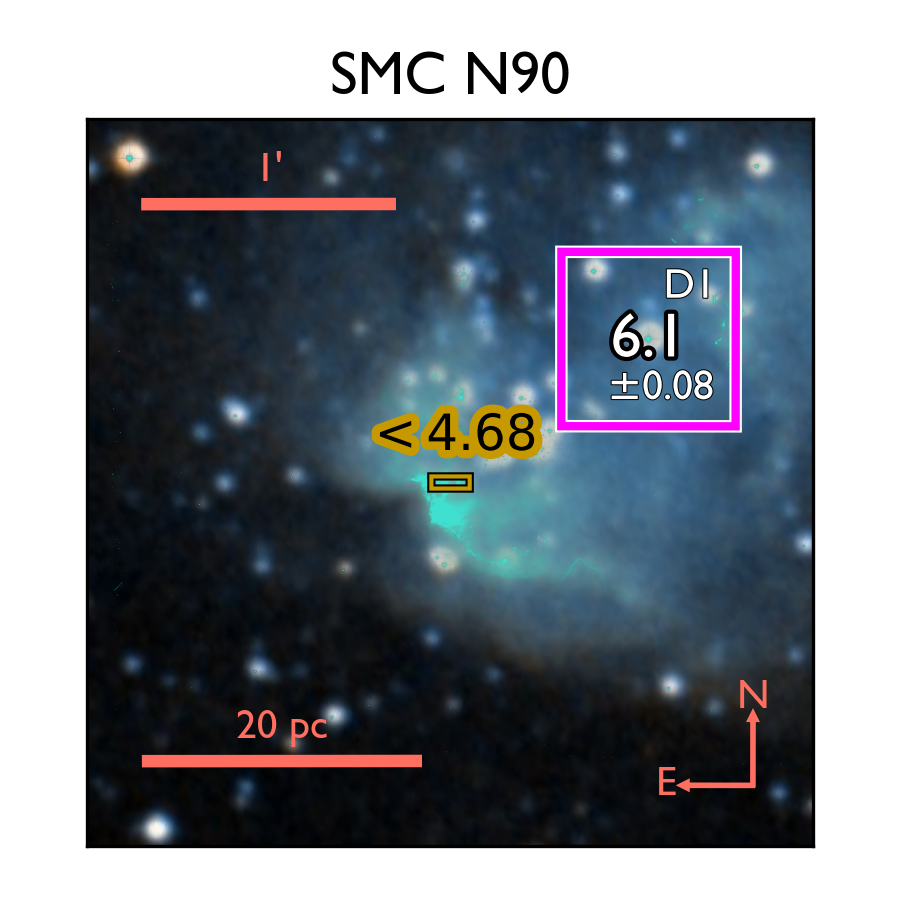}
    \caption{The spatial location of the SMC sightlines relative to the H\,\textsc{ii} regions. The large top panel shows the compact H\,\textsc{ii} region SMC N66A and neutral gas sightlines toward targets in the associated open cluster NGC 346. The bottom three panels show three H\,\textsc{ii} region slits each accompanied with one neutral gas sightline. Sightlines are indicated by squares (standard) and circles (where necessary to avoid obstructing data features). Slits are indicated by the rectangles. The large numbers within the squares or pointed at circles denote the 12 + log(Fe/H), while the uncertainty and the sightline ID are respectively labeled in the bottom right and the top right corner. The small golden rectangle denotes the slit locations and sizes used for the H\,\textsc{ii} region observations by \citet{toribio_san_cipriano_carbon_2017}. The H\,\textsc{ii} region Fe abundances are labeled above the slits. For the SMC N66A panel, the background image is an HST false-color image and the DSS-II \citep{mclean_status_2000} color image where targets are not captured. For the three bottom panels, the HST F656N (H-$\alpha$) image is overlaid in cyan atop the DSS-II color background. Background image credit: SMC N66A - ESA/Hubble and NASA, A. Nota, P. Massey, E. Sabbi, C. Murray, M. Zamani (ESA/Hubble), the DSS-II and GSC-II Consortia. SMC N81, SMC N88A, SMC N90 - HST; NASA/ESA, the DSS-II and GSC-II Consortia.}
    \label{fig:smc-regions-sightlines}
\end{figure*}

Figure~\ref{fig:smc-regions-sightlines} shows the locations of the SMC sightlines where the neutral gas Fe abundance is measured, and the locations and dimensions of the slits employed in emission spectra of SMC H\,\textsc{ii} regions. 12 + log(Fe/H) in neutral gas sightlines that probe all surrounding directions of SMC N66A are greater than 12 + log(Fe/H) in the compact H\,\textsc{ii} region itself. In the remaining three regions the 12 + log(Fe/H) is compared to that of the surrounding neutral ISM from one sightline, with SMC N88A having the smallest offset and SMC N90 having the largest offset. 

\begin{figure*}[ht!]
    \centering
    \includegraphics[width = 6in]{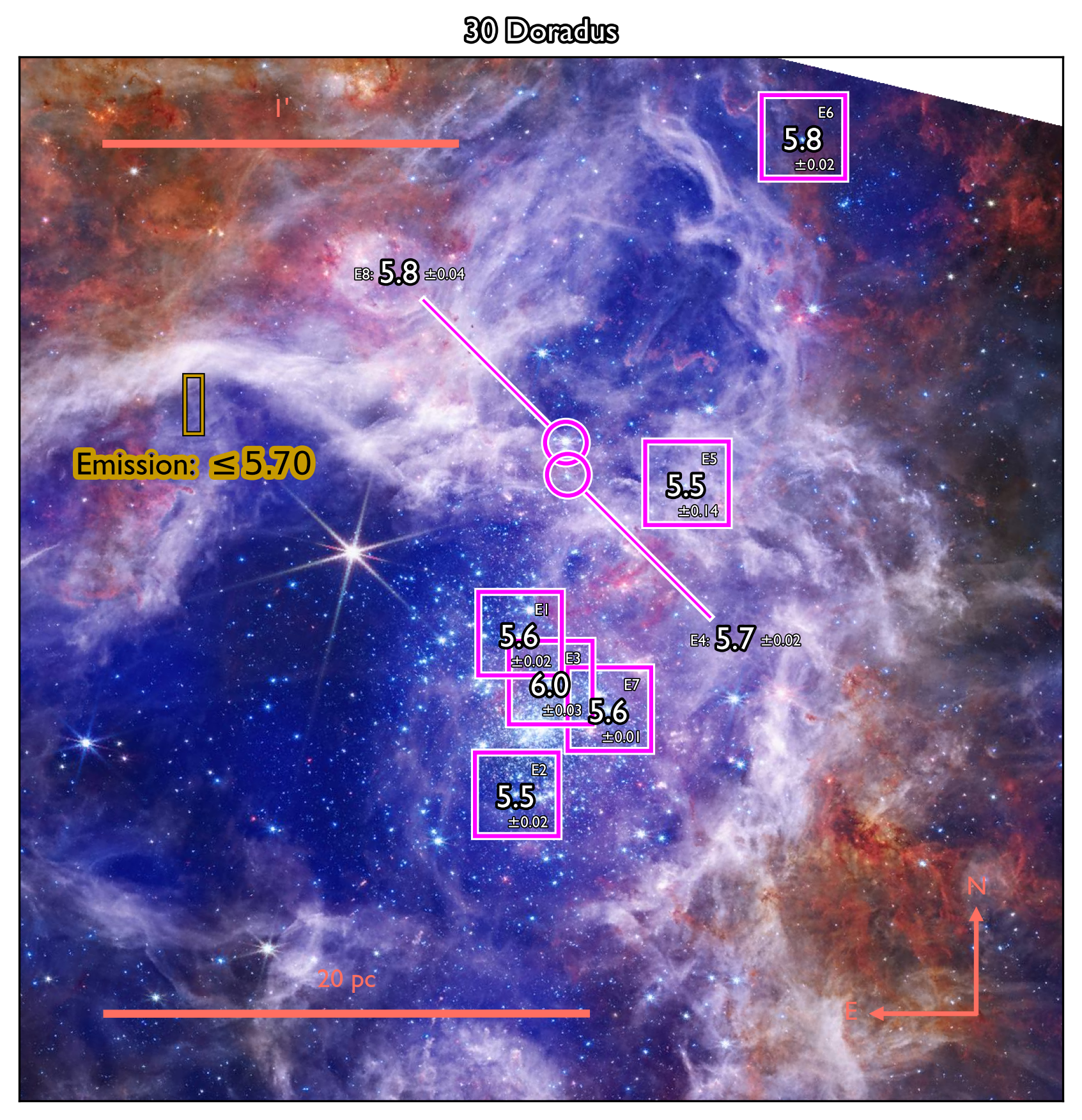}
    \caption{The spacial location of the LMC sightlines relative to 30 Doradus. Sightlines are indicated by squares (standard) and circles (where necessary to avoid obstructing data features). The slit is indicated by the rectangle. The large numbers within the squares or pointed at circles denote the 12 + log(Fe/H), while the uncertainty and the sightline ID are respectively labeled in the bottom right and the top right corner. The small golden rectangle denotes the slit locations and sizes used for the H\,\textsc{ii} region observations by \citet{peimbert_chemical_2003}. The H\,\textsc{ii} region Fe abundances are labeled above the slits. Background image credit: NASA. X-ray: NASA/CXC/Penn State Univ./L. Townsley et al.; IR: NASA/ESA/CSA/STScI/JWST ERO Production Team.}
    \label{fig:30-dor-sightlines}
\end{figure*}

\begin{figure*}[ht!]
    \centering
    \includegraphics[width = 6in]{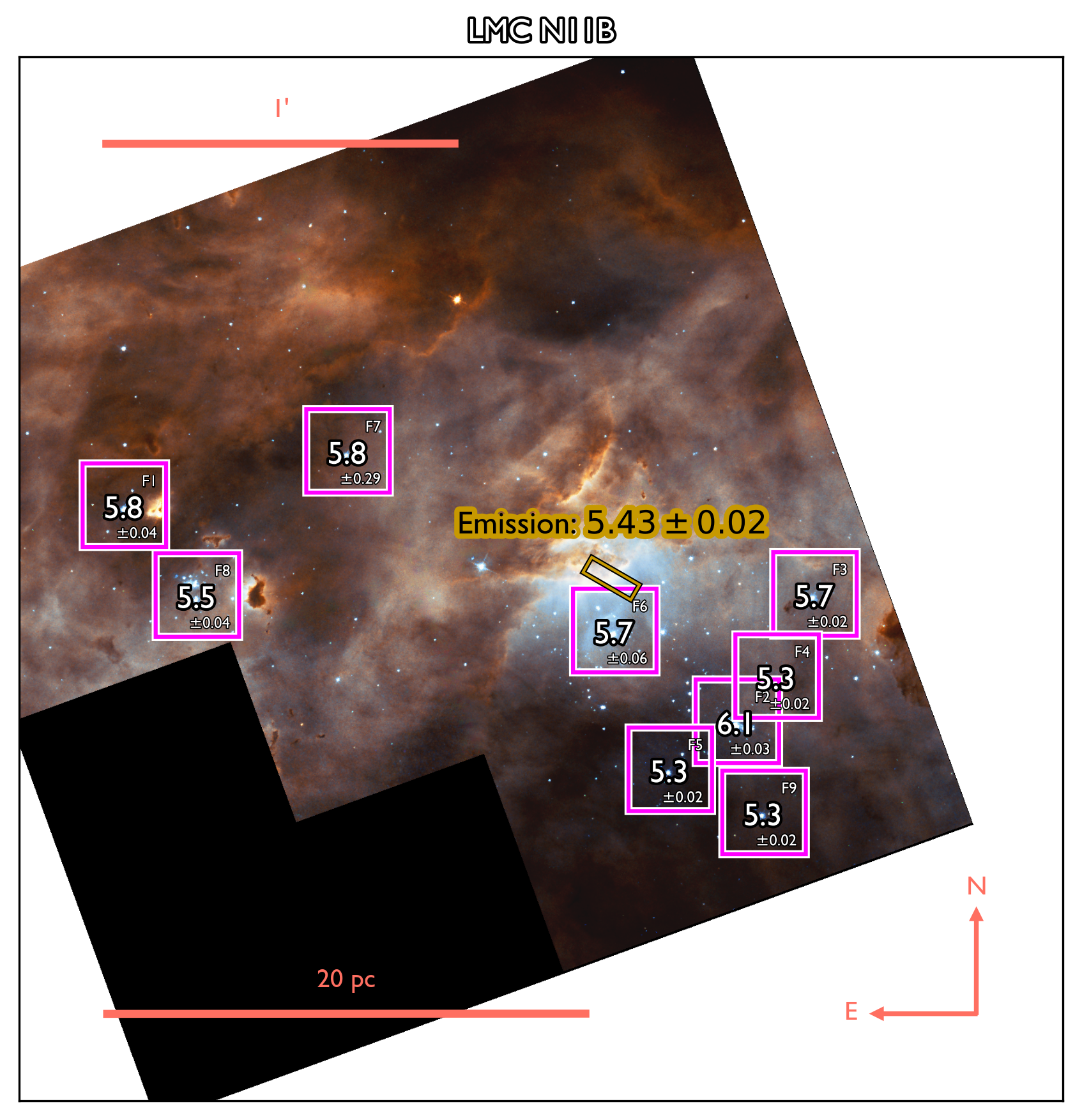}
    \caption{The spacial location of the LMC sightlines relative to LMC N11B. Sightlines are indicated by squares. The slit is indicated by the rectangle. The large numbers within the squares denote the 12 + log(Fe/H), while the uncertainty and the sightline ID are respectively labeled in the bottom right and the top right corner. The small golden rectangle denotes the slit locations and sizes used for the H\,\textsc{ii} region observations by \citet{toribio_san_cipriano_carbon_2017}. The H\,\textsc{ii} region Fe abundances are labeled above the slits. Background image credit: NASA/ESA and the Hubble Heritage Team (AURA/STScI)/HEIC.}
    \label{fig:n11b-sightlines}
\end{figure*}

Figure~\ref{fig:30-dor-sightlines} shows the parameters of the emission spectra slit toward LMC H\,\textsc{ii} region 30 Doradus observed by \citet{peimbert_chemical_2003}, and the associated neutral ISM sightlines toward the cluster NGC 2070. The sightlines are embedded within the extended H$\alpha$ emission from 30 Doradus, although they do not cover all directions that surround the H\,\textsc{ii} region. 12 + log(Fe/H) measured in neutral gas sightlines in general agree with the H\,\textsc{ii} Fe abundances. Figure~\ref{fig:n11b-sightlines} shows the parameters of the emission spectra slit toward H\,\textsc{ii} region LMC N11B observed by \citet{toribio_san_cipriano_carbon_2017}, and the associated neutral ISM sightlines. For this H\,\textsc{ii} region, the sightlines are embedded within the extended H$\alpha$ emission, and the direction coverage is excellent, with only the neutral ISM up north not comprehensively covered. 12 + log(Fe/H) measured in neutral gas sightlines marginally exceed the H\,\textsc{ii} Fe abundances ionization correction schemes. The offset observed between the ISM phases of LMC N11B is not as significant as the offset observed in SMC N66A. 

\subsection{Comparison with Previous Measurements} \label{compare-with-previous}

For the Voigt profile fitting determination of elemental column densities and the Ly$\alpha$ profile fitting determination of atomic hydrogen column densities, we compare our results to previous measurements. Using the method discuss in Section~\ref{voigt}, we measured all available transitions including S and Fe, and compare to previous works. Table~\ref{tab:comparison} details the comparison. 

\begin{table*}[t!]
\centering
\caption{Comparison with Previous Measurements}
\begin{tabular}{llllll}
\toprule
Target & H\,\textsc{ii} Region & Transition & This work & Reference & Literature value \\
\midrule
PGMW 3120 & LMC N11B & log(NH\,\textsc{i}) & 21.50 $\pm$ 0.02 & (1) & 21.48 $\pm$ 0.03 \\
PGMW 3120 & LMC N11B & log(NFe\,\textsc{ii}) & 15.240 $\pm$ 0.06 & (2) & 15.22 $\pm$ 0.05 \\
PGMW 3120 & LMC N11B & log(NS\,\textsc{ii}) & >15.828 & (2) & >15.70 \\
PGMW 3120 & LMC N11B & log(NCr\,\textsc{ii}) & 13.529 $\pm$ 0.073 & (2) & 13.62 $\pm$ 0.06 \\
PGMW 3120 & LMC N11B & log(NMg\,\textsc{ii}) & 16.149 $\pm$ 0.039 & (2) & 16.05 $\pm$ 0.12 \\
PGMW 3120 & LMC N11B & log(NNi\,\textsc{ii}) & 13.912 $\pm$ 0.044 & (2) & 14.00 $\pm$ 0.06 \\
PGMW 3120 & LMC N11B & log(NSi\,\textsc{ii}) & 15.904 $\pm$ 0.076 & (2) & 16.00 $\pm$ 0.26  \\
PGMW 3120 & LMC N11B & log(NZn\,\textsc{ii}) & 13.367 $\pm$ 0.043 & (2) & 13.34 $\pm$ 0.06 \\
BI 42 (PGMW 3223) & LMC N11B & log(NH\,\textsc{i}) & 21.46 $\pm$ 0.01 & (1) & 21.4 $\pm$ 0.06 \\
BI 42 (PGMW 3223) & LMC N11B & log(NFe\,\textsc{ii}) & 15.245 $\pm$ 0.037 & (2) & 15.24 $\pm$ 0.03 \\
BI 42 (PGMW 3223) & LMC N11B & log(NS\,\textsc{ii}) & >15.694 & (2) & >15.75 \\
BI 42 (PGMW 3223) & LMC N11B & log(NCr\,\textsc{ii}) & 13.508 $\pm$ 0.027 & (2) & 13.53 $\pm$ 0.06 \\
BI 42 (PGMW 3223) & LMC N11B & log(NCu\,\textsc{ii}) & <12.724 & (2) & <12.72 \\
BI 42 (PGMW 3223) & LMC N11B & log(NMg\,\textsc{ii}) & 16.124 $\pm$ 0.026 & (2) & 16.12 $\pm$ 0.11 \\
BI 42 (PGMW 3223) & LMC N11B & log(NNi\,\textsc{ii}) & 13.828 $\pm$ 0.019 & (2) & 13.93 $\pm$ 0.05 \\
BI 42 (PGMW 3223) & LMC N11B & log(NSi\,\textsc{ii}) & 15.949 $\pm$ 0.050 & (2) & 15.91 $\pm$ 0.17 \\
BI 42 (PGMW 3223) & LMC N11B & log(NZn\,\textsc{ii}) & 13.449 $\pm$ 0.031 & (2) & 13.42 $\pm$ 0.05 \\
HD 5980 & SMC N66A & log(NH\,\textsc{i}) & 21.06 $\pm$ 0.03 & (3) & 21.06 $\pm$ 0.04 \\
HD 5980 & SMC N66A & log(NFe\,\textsc{ii}) & 14.918 $\pm$ 0.03 & (4) & 14.88 $\pm$ 0.02 \\
HD 5980 & SMC N66A & log(NS\,\textsc{ii}) & 15.478 $\pm$ 0.004 & (4) & 15.57 $^{+0.04}_{-0.05}$ \\
HD 5980 & SMC N66A & log(NCr\,\textsc{ii}) & 12.994 $\pm$ 0.040 & (4) & 13.33 $^{+0.07}_{-0.08}$ \\
HD 5980 & SMC N66A & log(NMg\,\textsc{ii}) & 15.716 $\pm$ 0.057 & (4) & 15.82 $^{+0.09}_{-0.11}$ \\
HD 5980 & SMC N66A & log(NNi\,\textsc{ii}) & 13.481 $\pm$ 0.018 & (4) & 13.76 $^{+0.03}_{-0.04}$ \\
HD 5980 & SMC N66A & log(NSi\,\textsc{ii}) & 15.686 $\pm$ 0.015 & (4) & 15.57 $^{+0.04}_{-0.05}$ \\
HD 5980 & SMC N66A & log(NZn\,\textsc{ii}) & 12.674 $\pm$ 0.044 & (4) & 12.65 $^{+0.07}_{-0.08}$ \\
SK 80 & SMC N66A & log(NH\,\textsc{i}) & 20.98 $\pm$ 0.03 & (5) & 21.02 \\
\bottomrule
\end{tabular}
\smallskip
\label{tab:comparison}

\textbf{Sources:} (1) \citet{roman-duval_metal_2019} (2) \citet{roman-duval_metal_2021} (3) \citet{welty_interstellar_2012} with uncertainties provided by \citet{jenkins_interstellar_2017} (4) \citet{jenkins_interstellar_2017} derived from the apparent optical depths (AOD) (5) \citet{welty_interstellar_2012}

\end{table*}

The log(NH\,\textsc{i}) measurements agree with previous studies with discrepancies less than 0.1 dex. The column density measurements toward the LMC targets agree with \citet{roman-duval_metal_2021} deviating by less than 0.1 dex. The measurements for the S and Fe column densities toward the SMC targets show excellent agreements with \citet{jenkins_interstellar_2017} by $\sim0.1$ dex. Other measured transitions can agree less well with \citet{jenkins_interstellar_2017}, where the apparent optical depths (AOD) method is used instead of Voigt profile fitting. 

\section{Discussion} \label{discussion}

\subsection{Sulfur Un-depleted Within H\,\textsc{ii} Regions}

The measured sulfur (S) abundances in neutral gas are consistent with previous studies. We encounter two challenges identified in prior work: the saturation of strong S\,\textsc{ii} transitions and the presence of singly-ionized S in H\,\textsc{ii} regions \citep{jenkins_unified_2009}. The effect of unresolved saturation is present in LMC spectra \citep{roman-duval_metal_2021} and therefore the LMC neutral gas S abundances are labeled as lower limits in Figure~\ref{fig:s-abundances}. Similar to \citet{jenkins_interstellar_2017}, we notice the measured SMC neutral gas S abundance can exceed the ``intrinsic'' photospheric S abundance of the respective galaxies. \citet{jenkins_unified_2009} and \citet{jenkins_interstellar_2017} discuss that the elevated neutral gas S abundances is due to S\,\textsc{ii} present in the H\,\textsc{ii} regions being mis-attributed to the neutral gas. Since our sightlines border H\,\textsc{ii} regions, we expect the elevated S abundances observed in Figure~\ref{fig:s-abundances}. 

For the H\,\textsc{ii} regions in our sample, the S abundances can be slightly lower than the intrinsic abundance. This effect is unlikely to be caused by depletion into dust, but rather by an underestimation of the abundances of S$^{3+}$ and higher sulfur ionization states \citep{henry_curious_2012}. While S-bearing dust and molecules are known to be present in precursor dense clouds \citep{ruffle_sulphur_1999, martin-domenech_sulfur_2016}, these species are likely destroyed within H\,\textsc{ii} regions. 

\subsection{Iron Abundance Offset}

Comparing the iron (Fe) abundances in neighboring neutral and ionized components of the ISM in Figure~\ref{fig:fe-abundances} unexpectedly shows that Fe can be significantly \textbf{\textit{less}} abundant in gas-phase in the H\,\textsc{ii} regions. In SMC N66A, SMC N81, and SMC N90, the disagreement can exceed 0.5 dex. In LMC N11B, the difference is less extreme. This offset cannot be fully accounted for by inhomogeneities in the neutral ISM Fe abundance, as shown by the variations between sightlines. This is not the first example of Fe being observed more abundant in neutral-phase ISM than comparable ionized-phase ISM: \citet{james_classy_2026} observed neutral-phase Fe abundances higher than ionized-phase Fe abundances by $0.25\pm0.47$ dex ($0.66\pm0.61$ dex after ionization corrections) across a sample of 31 COS Legacy Archive Spectroscopic SurveY (CLASSY) galaxies at $z<0.2$, averaging over each galaxy. Our result shows that gas-phase Fe can be lower in the H\,\textsc{ii} regions by similar degrees on a parsec scale. In Sections~\ref{ic}~to~\ref{metal-poor-gas}, we begin by discussing mechanisms which are likely not driving this observation. The lower gas phase Fe abundances in the H\,\textsc{ii} regions likely results from the survival of Fe-bearing dust grains. This mechanism will be discussed in detail in Sections~\ref{survival}~to~\ref{growth}. 

\subsection{Abundance Offset Unlikely From Underestimated H\,\textsc{ii} Region Ionization Correction Factors} \label{ic}

We consider a scenario where the ionization corrections factors (ICFs) underestimate the H\,\textsc{ii} region Fe abundance. This hypothesis is motivated by three considerations: firstly, we observe large Fe abundance offsets between the neutral and the ionized gas. Secondly, systematic uncertainties in the ICF have a high impact because the inferred Fe\,\textsc{iv} abundance is often simultaneously the predominant ionization state (see Section~\ref{hii-abundances}). Finally, for SMC N88A where Fe\,\textsc{iv} is measured directly without relying on an ICF, no strong Fe abundance offset is observed. 

\begin{figure}[ht!]
    \centering
    \includegraphics[width = 3.4in]{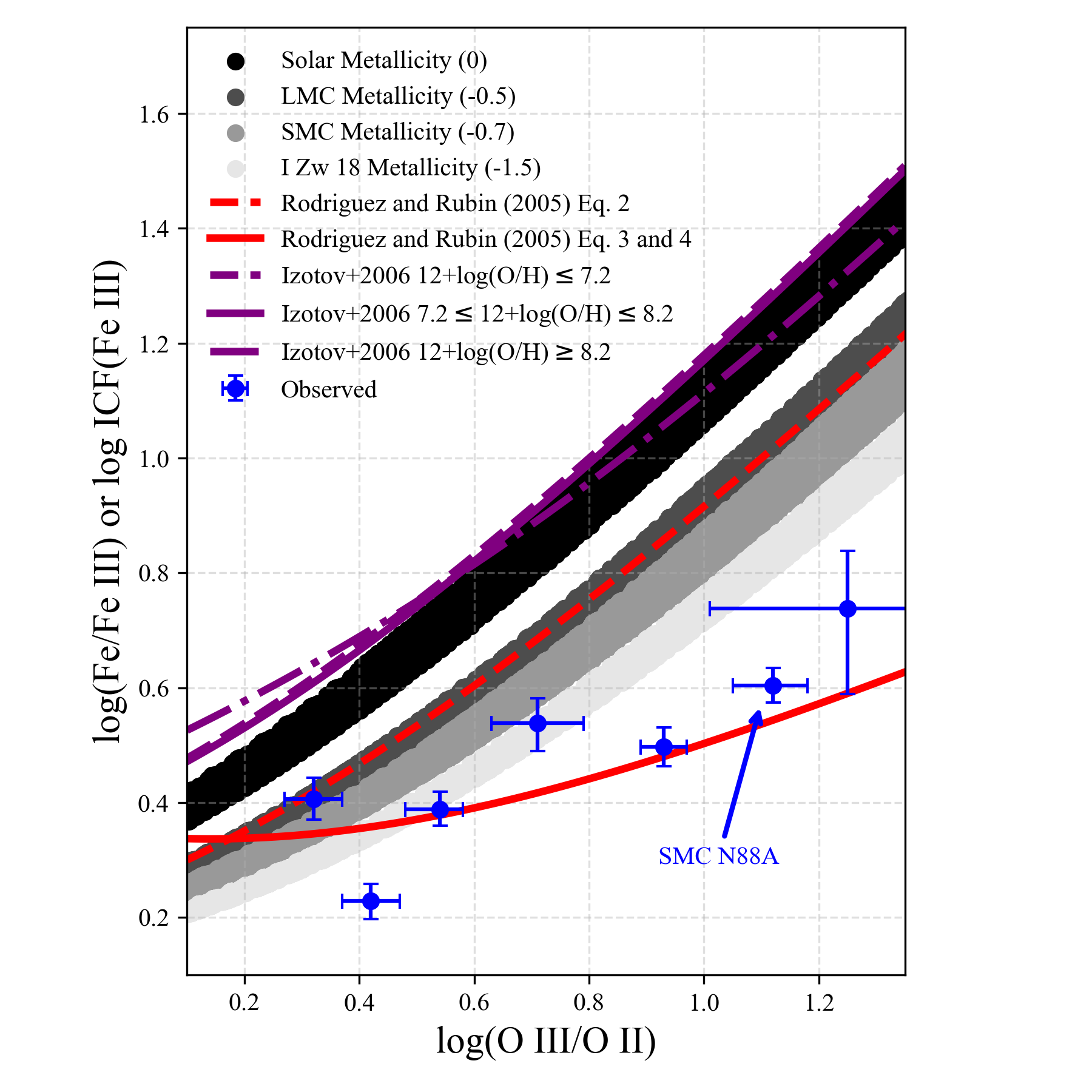}
    \caption{log(Fe/Fe\,\textsc{iii}), or log ICF(Fe\,\textsc{iii}) as a function of log(O\,\textsc{iii}/O\,\textsc{ii}) for all Cloudy input parameters (black and gray bands). In the Cloudy simulations, at different metallicities the ICF for Fe can be different by up to $\sim0.6$ dex, while at the same metallicity the ICF for Fe varies by up to $\sim0.1$ dex. Total Fe abundance decreases with metallicity. Seven H\,\textsc{ii} regions with observed Fe\,\textsc{iv} (including SMC N88A) are shown in blue. The ICFs of \citet{rodriguez_fe_2005} and \citep{izotov_chemical_2006} are respectively plotted in the red and purple lines. Eq.~2 of \citet{rodriguez_fe_2005} (red dashed line) and the equations of \citet{izotov_chemical_2006} (purple lines) are based on photoionization modeling. The ICF of Eq.~3~and~4 of \citet{rodriguez_fe_2005} (red solid line) is a least-squares fit to the observations. The Cloudy ICF agrees with the \citet{izotov_chemical_2006} ICF at Solar metallicity, but ICF variation with metallicity is higher in Cloudy. Eq.~2 of \citet{rodriguez_fe_2005} agrees well with the Cloudy ICF at the LMC metallicity (-0.5). H\,\textsc{ii} region Fe abundances adopted for this study employed Eq.~2 of \citet{rodriguez_fe_2005}.}
    \label{fig:ic}
\end{figure}

The H\,\textsc{ii} region Fe abundances accumulated for this study employ Eq.~2 of \citet{rodriguez_fe_2005} (red dashed line in Figure~\ref{fig:ic}), which computes Fe/Fe\,\textsc{iii} using the observed O\,\textsc{iii}/O\,\textsc{ii} ratio. This method does not explicitly account for the metallicity (12 + log(O/H)) of the H\,\textsc{ii} region. However, ICFs may vary in an H\,\textsc{ii} region with metallicity \citep{amayo_ionization_2021}, due to electron temperature differences \citep{shaver_galactic_1983, balser_metallicityelectron_2024}, and possibly variations in dust depletion mirroring trends observed in neutral gas \citep[e.g.][Figure 10]{roman-duval_metal_2022}. To gauge how the ICF changes with metallicity, we run the microphysics code Cloudy version C25 \citep{gunasekera_2025_2025}. In our inputs, we vary the metallicity of the medium from Solar, $-0.5$ (LMC), $-0.7$ (SMC), to $-1.5$ (I Zw 18); we vary the depletion factor (F$_*$) \citep{jenkins_unified_2009} from $-2.0$ (strongly depleted) to $0$ (undepleted) by steps of $0.1$. It needs to be pointed out that the depletion modeling in Cloudy C25 uses Milky Way neutral ISM parameters from \citet{jenkins_unified_2009} and does not account for depletion varying between ionization states. 

We list all the Cloudy parameters in Appendix \ref{cloudy}. We employ stellar SEDs from \citet{lanz_grid_2003} with varying ionization parameters. Using the star, we ionize a medium with varying log hydrogen density, metallicity, and depletion/dust abundance. For other settings, we use parameters recommended by Cloudy's documentation for an H\,\textsc{ii} region. For the resulting gas-phase composition of the simulated medium, we sample the abundance of each Fe and O ionization state at varying depths. We plot the resulting log(Fe/Fe\,\textsc{iii}), or log ICF(Fe\,\textsc{iii}), for all conditions with respect to the log(O\,\textsc{iii}/O\,\textsc{ii}) ratio in Figure~\ref{fig:ic}. As a comparison, we plot the ICFs of \citet{rodriguez_fe_2005} and \citep{izotov_chemical_2006}. 

The H\,\textsc{ii} region gas-phase Fe abundances in the current study are produced from Eq.~2 of \citet{rodriguez_fe_2005}. As shown in Figure~\ref{fig:ic}, Eq.~2 of \citet{rodriguez_fe_2005} in general overestimates the total Fe abundances compared to the observed sample. Adopting Eq.~3 and Eq.~4 of \citet{rodriguez_fe_2005} which is based on the observed sample increases the offset between the H\,\textsc{ii} regions and co-spatial neutral gas. Additionally, the variation introduced by lowering the metallicity in Cloudy leads to a lower total Fe abundance, also serving to increase the offset. Both findings point toward an overestimated ionized gas-phase Fe abundance and a higher offset of H\,\textsc{ii} region Fe abundances from neutral gas than shown in Figure~\ref{fig:fe-abundances}. Thus, it is unlikely that we can attribute the majority of the observed ionized-to-neutral offset to underestimated ICs alone. 

Various studies \citep[e.g.][]{dominguez-guzman_homogeneity_2022, mendez-delgado_gas-phase_2024} employ the \citet{rodriguez_fe_2005} ICFs after careful considerations, with Eq.~2 showing less discrepancies from observed values \citep[][also see Figure~\ref{fig:ic}]{mendez-delgado_gas-phase_2024}, and Eq.~3 and Eq.~4 providing reliable lower limit estimates \citep{dominguez-guzman_homogeneity_2022}. However, this method may still be subject to uncertainties in collision strengths \citep{rodriguez_fe_2005} and temperature inhomogeneities \citep{mendez-delgado_gas-phase_2024}. Such uncertainties are difficult to resolve given the current number of H\,\textsc{ii} regions with Fe\,\textsc{iv} measurements. Given the only neutral gas sightline co-spatial with SMC N88A is under-abundant in Fe relative to general SMC depletion trends, we recommend observing additional sightlines in this region to better constrain the Fe abundance offset between the H\,\textsc{ii} region and its surrounding neutral medium.

\subsection{Abundance Offset Unlikely From Sightline Blending}

Sightline blending is a common issue for absorption spectroscopy studies and may be present within SMC and LMC sightlines. While emissions from the H\,\textsc{ii} regions reflect the ionized ISM patches that are a few parsecs wide, absorption sightlines may be going through multiple clouds with distinct properties. Sightline-integrated abundances, which we measure in this study, can be biased by the blending of clouds with different metallicities and depletions along the sightline \citep{ritchey_distribution_2023, de_cia__2024}. It may be possible that measurements isolated to the parts of the sightlines directly adjacent to the H\,\textsc{ii} regions will yield higher Fe depletions similar to that observed in the H\,\textsc{ii} regions themselves. 

For the LMC, we do not expect significant sightline blending. The LMC is seen face on, thus the sightlines are perpendicular to the H\,\textsc{i} disk, and we will not observe metallicity or depletion gradients along directions parallel to the disk (e.g. \citet{hernandez_first_2021}). The thickness of the H\,\textsc{i} disk is measured to be $\sim100$ pc \citep{elmegreen_fractal_2001}, similar to the projected distance from the sightlines to the H\,\textsc{ii} region emissions. On the other hand, SMC sightlines have higher depths, thus it may be possible that they trace elemental abundances through extended, inhomogeneous regions of the neutral ISM. Another plausible scenario is that the targets are closer than the H\,\textsc{ii} regions, in which case the sightlines do not intersect the neutral ISM in the vicinity of the H\,\textsc{ii} regions.  

\citet{murray_galactic_2024} show that the SMC may host two superimposed ISM components with distinct radial velocities and a slight metallicity difference. The low velocity component and the high velocity component are respectively in the foreground across approximately half the disk of the SMC. The mean Fe/H for the stars in each component is different by $<0.1$ dex. We explored the possibility that the two components described in \citet{murray_galactic_2024} cause sightline blending in our data, and consequently produce the observed offset between Fe depletion in the neutral gas and that in the ionized gas. We cross-matched the location and radial velocities of our SMC H\,\textsc{i}/H\,\textsc{ii} sample with the superimposed ISM components discussed in \citet{murray_galactic_2024}. We find that SMC N66A, SMC N81, and SMC N90 reside within the component that is in the foreground ISM component, while SMC N88A resides where the two components are blended. Using relative velocities derived from jointly fitting the S\,\textsc{ii} $\lambda\lambda$ 1250, 1253 {Å} absorption profiles, and matching to the position-velocity-foreground/background relation presented in \citet{murray_galactic_2024}, we find that likely all sightlines path through the neutral ISM adjacent to the H\,\textsc{ii} regions. This is supported the high observed N(H\,\textsc{i})$_{\text{MC}}$. Additionally, the observed super-intrinsic 12 + log(S\,\textsc{ii}/H) can indicate tracing S\,\textsc{ii} from ionized gas along the sightlines, implying they intersect the neighboring neutral ISM as well. 

Most of the absorption lines within our sample did not show signs of multiple components with distinct relative velocities. For the E230M spectra of the HD 5980 (A12, AzV 229, SK 78) sightline, we performed a test fit of the Fe and Zn absorption lines, assuming two components with guess radial velocities adopted from \citet{murray_galactic_2024}. We did not find a significant difference ($>0.1$ dex) in the total Fe, Zn abundances derived with this method, nor was there a significant difference in depletion traced by Fe/Zn. Since the saturated Ly$\alpha$ absorption line do not allow accurately fitting more than one Magellanic Clouds atomic hydrogen component, we did not perform a comparison of Fe/H or Zn/H in both components. 

For our adopted neutral ISM observations, the majority of the targets are luminous Wolf-Rayet or OB stars (e.g. ULLYSES \citep{roman-duval_uv_2025}, METAL \citep{roman-duval_metal_2019}). It is reasonable to concern that the observed sightlines are translucent low $A_v$ sightlines; dust abundance along the sightlines are low (but high elsewhere near the H\,\textsc{ii} regions); and we would expect a lower depletion compared to nearby H\,\textsc{ii} regions. This is unlikely the case given understood Magellanic Clouds neutral ISM inhomogeneities \citet{tchernyshyov_elemental_2015, jenkins_interstellar_2017, roman-duval_metal_2021}. 

\subsection{Abundance Offset Unlikely From Cold, Metal-poor Gas Accretion} \label{metal-poor-gas}

Previous studies have reported metal-poor H\,\textsc{ii} regions through comparing to other emission abundances within the same system: \citet{sanchez_almeida_localized_2015} detected strong inhomogeneities of metallicity in ten extremely metal-poor galaxies (XMPs), where emissions from the star-forming regions show a 12 + log(O/H) that is 1 dex lower than emissions from the rest of the galaxies; \citet{lagos_detecting_2018} found a low-metallicity region in an H\,\textsc{ii} region of dwarf galaxy UM 461. Both studies report their observations to be consistent with the recent accretion of cold, metal-poor gas from a cosmic cloud or the galaxy's own neutral ISM. In this study, we observe the parsec-scale neighboring neutral ISM and do not identify such low metallicity clouds. It is unlikely that such low-metallicity clouds preferentially mix with H\,\textsc{ii} regions at a parsec-scale. 

\subsection{Abundance Offset Results From Fe Dust Survival in H\,\textsc{ii} Regions} \label{survival}

The offset in the gas-phase Fe abundances may reflect a difference in the depleted Fe fraction between the cold neutral medium (CNM) and the H\,\textsc{ii} regions. Although grain growth via accretion is suppressed in H\,\textsc{ii} region environments (see Section~\ref{growth}), the dense atomic or molecular clouds from which H\,\textsc{ii} regions formed can be highly depleted in Fe. Observing nearly fully Fe depleted H\,\textsc{ii} regions with an equivalent or stronger Fe depletion compared to co-located neutral gas indicate grain survival on Myr timescales. 

\subsubsection{Slow Grain Destruction by Supernovae Shocks in H\,\textsc{ii} Regions} \label{sne-shocks}

On a galaxy-wide scale, grains are primarily destroyed via sputtering in supernovae shocks \citep{draine_physics_1979, mckee_dust_1989, zhukovska_modeling_2016}. Within our sample, SMC N66A is co-located with SNR 0057-7226, a supernova remnant expanding from the foreground into the extended H\,\textsc{ii} region N66 \citep{danforth_far-ultraviolet_2003}. SMC N66A remains highly depleted in Fe. Therefore given the current observations, we cannot distinguish between the following scenarios: the SNR has not reached the compact H\,\textsc{ii} region; grains were ineffectively sputtered; or grains were destroyed non-locally and Fe was not restored to the H\,\textsc{ii} region gas. Correlating with the locations and sizes of supernova remnants (SNRs) in the Magellanic Clouds listed in \citet{temim_dust_2015}, we find that the other five H\,\textsc{ii} regions do not intersect known SNRs. We do not expect thermal sputtering of Fe dust within these H\,\textsc{ii} regions since this process is only effective at temperatures beyond $10^6$ K \citep{draine_physics_2011}, much higher than that of the H\,\textsc{ii} regions in our sample \citep{dominguez-guzman_homogeneity_2022}. Therefore, Fe dust is unlikely destroyed within supernovae shocks. 

\subsubsection{Slow Photodestruction in H\,\textsc{ii} regions} \label{photodestruction}

Grains can be destroyed in H\,\textsc{ii} regions given evidence of the hard radiation fields causing the photodestruction of polycyclic aromatic hydrocarbons (PAHs) \citep{chastenet_polycyclic_2019, chown_pdrs4all_2024, sutter_fraction_2024}. The destruction of PAHs ($a<1$ nm) shows that very small grains are vulnerable to UV radiation. However, Fe-bearing grains that built up via gas-phase accretion are likely larger because efficient accretion rapidly grows grains beyond 1 nm \citep{hirashita_dust_2012}. Consequently, it is expected that the majority of these grains should survive. 

The H\,\textsc{ii} regions included for this study are depleted by $>95\%$. Thus, photodestruction would have caused at most a few percent of Fe dust destruction over the age of the H\,\textsc{ii} regions of several Myrs. This scenario implies a photodestruction timescale lower limit of $\gtrsim100$ Myr. This lower limit is not particularly constraining, nor does it align with understood photodestruction efficiencies. The destruction of PAHs by UV radiation occurs on timescales of a few years \citep{allain_photodestruction_1996}. Therefore, photodestruction is unlikely to have played a significant role in the observed Fe depletion. 

\subsubsection{Insubstantial Astration in H\,\textsc{ii} regions} \label{astration}

In H\,\textsc{ii} regions, grains can be destroyed via astration, where they are incorporated into forming stars. While it is typically assumed that astration consumes gas and dust non-preferentially, recent models suggest gas is preferentially accreted while dust grains are more efficiently radiatively evacuated from the stellar vicinity \citep{soliman_dust-evacuated_2024}. Therefore, no more than a few percent of Fe dust will ultimately become stellar material. If gas and dust are removed non-preferentially, since the elements locked into stars during astration do not return to the gas phase, the observed gas-phase Fe depletion will be unaffected by this mechanism.

\subsection{Constraining the Mixing Timescale between H\,\textsc{ii} Region Gas and the Diffuse ISM} \label{mixing}

The non-destruction of Fe dust maintained the Fe-depleted state of H\,\textsc{ii} region gas. However, mixing between Fe-depleted H\,\textsc{ii} regions and their surrounding, less-depleted diffuse ISM will reduce the Fe depletion within the H\,\textsc{ii} regions themselves. The Fe depletion cannot increase again in H,\textsc{ii} regions because grain growth is ineffective at high temperatures \citep{bossion_accurate_2024}. Therefore, we can constrain the timescales of mixing between the H\,\textsc{ii} regions and diffuse gas by assuming that Fe is fully depleted at the end of the molecular cloud phase\footnote{As we discuss later, this requires sub-Myr growth timescales.}, and mixing reduced the Fe depletions to the currently observed H\,\textsc{ii} region abundances. By incorporating assumptions of the H\,\textsc{ii} regions' age, we obtain lower limits to the mixing timescale ($\tau_{\rm m}$) between the H\,\textsc{ii} region and surrounding diffuse ISM. We define the mixing timescale $\tau_{\rm m}$ with the integral 

\begin{equation}
\label{eqn:mixing}
    \Delta f_{\rm Fe,\,gas}=\int^{\tau_{\rm H\,II}}_{0} \frac{f_{\rm Fe,\,gas,\,H\,I}-f_{\rm Fe,\,gas}(t)}{\tau_{\rm m}} \, dt,
\end{equation}

where $\Delta f_{\rm Fe,\,gas}$ is the observed change in the H\,\textsc{ii} region gas-phase Fe fraction, $f_{\rm Fe,\,gas,\,H\,I}$ is the gas-phase Fe fraction observed in the surrounding neutral ISM, $f_{\rm Fe,\,gas}(t)$ is the Fe depletion at time $t$, and $\tau_{\rm H\,II}$ is the lifetime of the H\,\textsc{ii} region. We assume $\tau_{\rm H\,II}\sim1$ Myr for LMC N11B \citep{walborn_hstfos_1999} and $\tau_{\rm H\,II}\sim3$ Myr for SMC N66A \citep{heydari-malayeri_very_2010}. Using the observed depletion for both H\,\textsc{ii} regions and surrounding neutral gas, we obtain a mixing timescale lower limit of $\tau_{\rm m}\gtrsim1-10$ Myr. This derived lower limit is significantly lower than analytically derived mixing timescales of $\sim$40 Myr \citep{roy_dispersal_1995}, 100--300 Myr \citep{krumholz_metallicity_2018}, and remains consistent with other observational lower bounds of $\sim$10 Myr \citep{kreckel_measuring_2020}. 

\subsection{Grain Growth by Accretion in the CNM and Molecular Clouds} \label{growth}

Gas-phase species collide with dust grains in the ISM. Under certain conditions, they may be incorporated into grain material, resulting in grain growth over time. This accretion process which results in the observed depletion in the ISM \citep[e.g.][]{jenkins_unified_2009, roman-duval_metal_2021} is considered a primary growth mechanism of dust in local galaxies \citep{zhukovska_evolution_2008, dwek_iron_2016}. Gas-dust accretion is believed to occur primarily in cold, dense phases of the ISM such as the cold neutral medium (CNM) and molecular clouds \citep{jenkins_unified_2009, tchernyshyov_elemental_2015, jenkins_interstellar_2017, roman-duval_metal_2022}, but not in H\,\textsc{ii} regions. This is also supported by molecular dynamic simulations which find the sticking efficiency of H$_2$ and CO are effectively zero for a gas temperature above 1000 K \citep{bossion_accurate_2024}. 

Although grain growth via accretion is suppressed in H\,\textsc{ii} region environments, an H\,\textsc{ii} region will inherit strong Fe depletion from the dense atomic or molecular clouds from which it formed. This inherited depletion is likely stronger than that in the surrounding diffuse neutral ISM due to the higher densities of preceding clouds facilitating more efficient accretion of gas-phase Fe onto dust grains. Direct UV absorption studies of dense clouds are difficult to perform, but \citet{welty_hd_2020} show that translucent cloud HD 62542 exhibits a strong $-2.93$ dex Fe depletion in the core. Beyond direct observations of clouds, the density-depletion relationship is well-supported both observationally and theoretically. Elemental depletions, including Fe, correlate with hydrogen column density \citep{jenkins_unified_2009, tchernyshyov_elemental_2015, roman-duval_metal_2021, roman-duval_metal_2022, hamanowicz_metal-z_2024}. Furthermore, \citet{clark_quest_2023} show that the deprojected surface density of hydrogen is positively correlated to the dust-to-gas ratio derived from observed dust emission. Finally, numerical hydrodynamical simulations of giant molecular clouds show that silicon and iron depletion increase with cloud density \citep{zhukovska_modeling_2016, zhukovska_iron_2018}. 

\begin{figure*}[ht!]
    \centering
    \includegraphics[width = 6in]{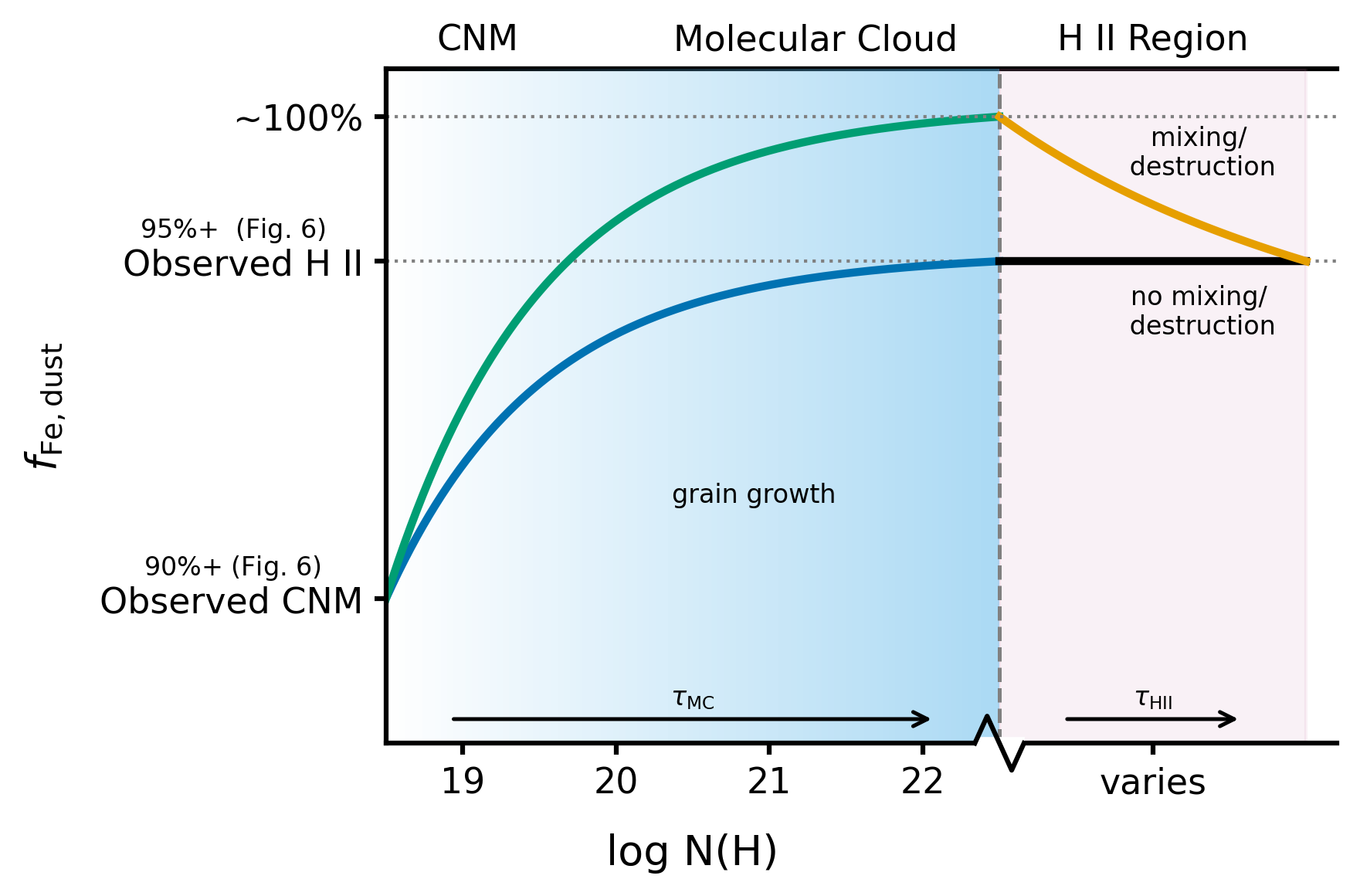}
    \caption{A schematic of the evolution of the Fe fraction depleted into dust grains ($f_{\rm Fe,\,dust}$). The vertical axis denotes the dust depletion fraction, while the horizontal axis indicates the evolutionary stage of the local ISM, labeled with the hydrogen column density. The horizontal arrow labels the general trend of the dynamic density evolution within the lifetime of the molecular cloud $\tau_{\rm MC}$. The axis break denote the transition from a molecular cloud to an H\,\textsc{ii} region. Within H\,\textsc{ii} regions, grain growth cannot continue, thus the depleted Fe fraction doesn't increase. Grain destruction/mixing may cause the depleted Fe fraction to level off or decrease, depending on the maximum depleted Fe fraction. These processes are discussed from Section~\ref{survival} to Section~\ref{mixing}. The maximum depleted Fe fraction depends on the grain growth rate and growth timescales, which are discussed in Section~\ref{growth}.}
    \label{fig:schematic}
\end{figure*}

With the above analysis, we illustrate the evolution of the Fe fraction in dust $f_{\rm Fe,\,dust}$ in Figure~\ref{fig:schematic}. As a gas parcel cools and its hydrogen density increases to form a molecular cloud, $f_{\rm Fe,\,dust}$ increases simultaneously via grain growth by accretion. Once the molecular cloud evolves into an H\,\textsc{ii} region, grain growth is suppressed. This allows us to observationally constrain the grain growth timescale $\tau_{\rm g}$ \citep{dwek_evolution_1998, hirashita_dust--gas_1999, zhukovska_evolution_2008, zhukovska_modeling_2016, choban_galactic_2022} using the initial and final $f_{\rm Fe,\,dust}$ inferred for each H\,\textsc{ii} region in our sample summarized in Table~\ref{tab:f-fe-dust}. Specifically, we take the initial $f_{\rm Fe,\,dust}$ of the parcel to be the observed $f_{\rm Fe,\,dust}$ in the diffuse ISM surrounding each H\,\textsc{ii} region, assuming the parcel was initially chemically homogeneous with its surroundings; because grain growth continues in the diffuse ISM, this observed value is an upper limit on the parcel's true initial $f_{\rm Fe,\,dust}$ (with overestimation $\lesssim1\%$ for MRN grains and $\sim5\%$ for iron nanoparticles, discussed in Appendix~\ref{timescale}). We take the final $f_{\rm Fe,\,dust}$ to be the observed value within the H\,\textsc{ii} region itself; because mixing and grain destruction may reduce $f_{\rm Fe,\,dust}$ after the molecular cloud phase ends, this observed value is a lower limit on the true final $f_{\rm Fe,\,dust}$ reached in the cloud. 

\begin{table}[ht!]
\centering
\caption{Adopted $f_{\rm Fe,\,dust}$ for Growth timescale Estimation}
\begin{tabular}{lll}
\toprule
H\,\textsc{ii} region & Initial & Final \\
\midrule
30 Doradus & 0.976 $^{+0.003}_{-0.003}$ & $\geq$0.976 \\
LMC N11B & 0.980 $^{+0.004}_{-0.005}$ & 0.987 $^{+0.001}_{-0.001}$ \\
SMC N66A & 0.885 $^{+0.014}_{-0.016}$ & 0.977 $^{+0.001}_{-0.001}$ \\
SMC N81 & 0.830 $^{+0.015}_{-0.017}$ & 0.961 $^{+0.002}_{-0.002}$ \\
SMC N88A & 0.947 $^{+0.003}_{-0.003}$ & 0.950 $^{+0.003}_{-0.004}$ \\
SMC N90 & 0.850 $^{+0.024}_{-0.028}$ & $>$0.994 \\
\bottomrule
\end{tabular}
\smallskip
\label{tab:f-fe-dust}
\end{table}

We adopt a molecular cloud lifetime of 10--20 Myr \citep{hartmann_rapid_2001, vazquez-semadeni_molecular_2007} for all six H\,\textsc{ii} regions. With the above assumptions, we constrain the upper limit to the grain growth timescale $\tau_{\rm g}$ to vary between 5--50 Myr across our sample. The derivations are discussed in Appendix~\ref{timescale}. 

These estimated timescales are consistent with theoretical expectations, using the expression from \citet{zhukovska_evolution_2008, zhukovska_modeling_2016, zhukovska_iron_2018, choban_galactic_2022}. At a fixed $n(\rm H)=10^3$ cm$^{-3}$ and $T=20$ K, regular-sized \citep[quasi-MRN,][]{mathis_size_1977} metallic iron grains have $\tau_{\rm g}=41.3$ Myr and $\tau_{\rm g}=15.4$ Myr respectively for SMC and LMC metallicity, while iron nanoparticles \citep{zhukovska_iron_2018} have $\tau_{\rm g}=4.13$ Myr and $\tau_{\rm g}=1.54$ Myr respectively for SMC and LMC metallicity (detailed in Appendix~\ref{timescale}). MRN-sized metallic iron grains will grow in molecular clouds to approximately equal to the Fe depletion observed in the H\,\textsc{ii} regions (blue-black trajectory in Figure~\ref{fig:schematic}), while iron nanoparticles will grow to approximately $99.9\%$ (SMC metallicity) and $99.99\%$ (LMC metallicity), with depletion subsequently reduced to the observed H\,\textsc{ii} region levels via mixing and/or grain destruction (green-orange trajectory in Figure~\ref{fig:schematic}). Because the true Fe depletion at the end of the molecular cloud phase is unknown, current observations cannot distinguish between these grain models. Future work may better constrain models by placing stronger priors on cloud age, local mixing efficiency, and ongoing grain processing (e.g., grain growth and destruction in the surrounding diffuse ISM); adopting simulations with dynamic temperature and density; and reducing uncertainties caused by molecular hydrogen and ICFs.

Nevertheless, the observation of co-located H\,\textsc{ii} regions and neutral ISM may be promising for constraining the nature of Fe-bearing grains, if more locally-specific assumptions are adopted. Grain evolution models employing quasi-MRN metallic iron grains (with or without Coulomb enhancement) underpredict the observed Fe depletion in the diffuse neutral ISM \citep{zhukovska_evolution_2008}. While introducing iron nanoparticles enhances the accretion rate and reproduces the observed Fe depletions \citep{zhukovska_modeling_2016, choban_galactic_2022}, such nanoparticle populations are not favored by recent grain evolution simulations where the multiphase ISM is explicitly resolved \citep{choban_ashes_2026}. If Fe-bearing grains exist in quasi-MRN size distributions, the strong Fe depletion observed in the ISM of local galaxies \citep{jenkins_unified_2009, jenkins_interstellar_2017, roman-duval_metal_2021} may instead be attributed to metallic iron grains being highly resistant to grain destruction processes \citep{zhukovska_iron_2018, choban_ashes_2026}. While the current observations do not provide sufficient constraints on the size distribution of Fe-bearing grains, this framework shows that comparing gas-phase Fe abundances between H\,\textsc{ii} regions and co-located neutral ISM provides an indirect means to probe Fe depletion within dense molecular clouds, which are otherwise difficult to access via UV absorption spectroscopy. 

\section{Conclusion} \label{conclusion}

We compared the S and Fe abundances between H\,\textsc{ii} regions and neutral gas sightlines separated by parsec-scale distances in the Magellanic Clouds. The gas-phase S abundances compare as expected, but we find that the gas-phase Fe abundances can be lower in the H\,\textsc{ii} regions by $\sim0.5$ dex or more, as shown in Figure \ref{fig:fe-abundances}. With small physical separations between the compared environments, the under-abundance of gas-phase Fe in the H\,\textsc{ii} regions likely results from grain growth in the precursor molecular clouds and subsequent Fe dust survival. The key findings of this study can be summarized as follows: 

1. We measured the neutral gas elemental abundances in sightlines near six SMC and LMC H\,\textsc{ii} regions. As shown in Figure \ref{fig:fe-abundances}, four out of the six H\,\textsc{ii} regions have lower observed gas-phase 12 + log(Fe/H) compared to neutral gas in vicinity ($\sim0.5$ dex). Two of the Fe-deficient H\,\textsc{ii} regions, LMC N11B and SMC N66A, have 9 and 13 neutral ISM abundance tracers respectively that cover all directions surrounding the H\,\textsc{ii} regions. Given the proximity of the compared neutral gas, this result further substantiates that Fe is depleted in H\,\textsc{ii} regions. We show that Fe depletion within the H\,\textsc{ii} regions can be stronger than surrounding neutral gas. 

2. Determining Fe abundances in H\,\textsc{ii} regions relies heavily on ionization corrections. Current photoionization modeling-based ionization correction methods carry systematic uncertainties, but have generally been shown to overestimate the total Fe abundance in ionized gas \citep{rodriguez_fe_2005, mendez-delgado_gas-phase_2024}. Using Cloudy C25 \citep{gunasekera_2025_2025}, we find that lower intrinsic metallicity may cause a further overestimation of the total Fe abundance (Figure~\ref{fig:ic}). Therefore, errors in ionization corrections are unlikely to account for the observed underabundance of gas-phase Fe in H\,\textsc{ii} regions. Nonetheless, the impact of intrinsic metallicity on Fe ionization corrections in H\,\textsc{ii} regions should be revisited. 

3. Compared to the Fe gas-phase abundances in the neighboring neutral ISM, the deficiency of gas-phase Fe in the H\,\textsc{ii} regions is likely an effect of dust depletion. As illustrated in Figure~\ref{fig:schematic}, large amounts of gas-phase Fe is accreted into dust in the dense atomic or molecular clouds prior to the formation of the H\,\textsc{ii} regions, while those Fe-bearing grains survive subsequent destruction. We show that observing H\,\textsc{ii} regions allows probing Fe depletion within dense molecular clouds that are difficult to observe via absorption spectroscopy; comparing the Fe depletion in the H\,\textsc{ii} regions with that of the surrounding diffuse ISM allows characterizing both the mixing timescales of neutral and ionized gas and grain growth timescales within the precursor molecular clouds. While our current constraints are broad due to uncertainties in the local environment evolutionary history, future studies utilizing assumptions based on local observations will provide the tighter constraints to these timescales. 

4. The observed sulfur (S) abundances in neighboring neutral and ionized patches of the ISM shown in Figure~\ref{fig:s-abundances} are as expected. In the neutral gas, we observe super-intrinsic S abundances due to sightlines intersecting H\,\textsc{ii} regions. In the ionized gas, the lower S abundances are likely not an effect of depletion into dust. 

Our comparison shows that the offset in Fe depletion is greater in the SMC than in the LMC, driven by a lower baseline Fe depletion in the low-metallicity SMC neutral ISM. However, three of the four SMC H\,\textsc{ii} regions correspond to only a single neutral gas sightline each. Expanding this sample by observing additional neutral gas sightlines near SMC N81, SMC N88A, and SMC N90 would be highly valuable. SMC N88A is a particularly compelling case, with Fe\,\textsc{iv} directly detected, providing Fe abundances that are unaffected by ionization correction uncertainties. We also recommend observing H\,\textsc{ii} regions with existing $12 + \log(\rm Fe/H)$ measurements that lack neighboring archival neutral gas sightlines. 

Given this result indicates stronger Fe depletion in the H\,\textsc{ii} regions, we may be able to observe additional dust signatures in absorption or emission within these environments. One potential future experiment is to search for enhanced silicate extinction in the H\,\textsc{ii} regions. A fraction of the depleted gas-phase Fe may reside within silicates, which can be detected via the 9.7 $\mu$m and the 18 $\mu$m features in absorption spectra. The relationship between elemental depletions in neutral gas and silicate absorption has previously been established for the Milky Way \citep{decleir_first_2025, zeegers_investigating_2025} and observed within the Magellanic Clouds \citep{gordon_first_2024}. If the depleted Fe is in the form of silicates (or metallic iron included within silicates \citep{zhukovska_iron_2018}), the observed silicate extinction toward targets within the H\,\textsc{ii} regions may be stronger than anticipated based on trends established by the aforementioned observations. 

\vspace{+6pt}


\begin{acknowledgments}

Y. Q. L., J. K. W., and K. T. gratefully acknowledge the support from NSF-CAREER 2044303. 

Y. Q. L. thanks M. McQuinn, T. R. Quinn, and R. Barnes for providing helpful comments on a draft version of this article. Discussions with K. D. Gordon on iron dust were important to the final ideation of this study. A. De Cia gave helpful advice on element selection and $\alpha$-element enhancement. Y. Q. L. thanks M. R. Morris for discussions on iron dust, iron in extremely ionized environments, and much encouragement. B. Choi provided great help with software setup and many comments that helped improve this manuscript. B. Benda provided great comments for the figures. 

Data analyzed in this work are obtained from the Mikulski Archive for Space Telescopes (MAST) archive, maintained by the Space Telescope Science Institute (STScI), at \dataset[doi:10.17909/541z-0x45]{\doi{10.17909/541z-0x45}}. 

This research has made use of "Aladin sky atlas" developed at CDS, Strasbourg Observatory, France \citep{bonnarel_aladin_2000}. This research has made use of the SIMBAD database, operated at CDS, Strasbourg, France \citep{wenger_simbad_2000}. This research has made use of the NumPy \citep{van_der_walt_numpy_2011, harris_array_2020}, SciPy \citep{virtanen_scipy_2020}, Matplotlib \citep{hunter_matplotlib_2007}, Pandas \citep{mckinney_data_2010, the_pandas_development_team_pandas-devpandas_2026}, and emcee \citep{foreman-mackey_emcee_2013} Python packages. This work made use of Astropy: \footnote{https://www.astropy.org} a community-developed core Python package and an ecosystem of tools and resources for astronomy \citep{astropy_collaboration_astropy_2022}. This work made use of The Potsdam Wolf-Rayet Models (PoWR) \citep{sander_galactic_2012, todt_potsdam_2015, hainich_powr_2019}. This research has made use of Linetools, a package for the analysis of 1d astronomical spectra, especially quasar and galaxy spectra \citep{prochaska_linetoolslinetools_2016}. This work has made use of pyigm, an astropy-affiliated package developed to provide software useful for research on the Intergalactic Medium (IGM) \citep{prochaska_pyigmpyigm_2017}. This work has made use of the Veeper, a Voigt profile fitter for CGM absorption lines \citep{burchett_veeper_2024}. This work made use of calculations performed with version C25 of Cloudy (last described by \citet{gunasekera_2025_2025}). This research has made use of the hips2fits, a tool developed at CDS, Strasbourg, France aiming at extracting FITS images from HiPS sky maps with respect to a WCS. 

Facilities: HST (COS, STIS), VLT (UVES)

\end{acknowledgments}

\bibliography{floatformorrow.bib}

\appendix

\section{Cloudy Parameters Employed for Rudimentary Ionization Corrections Modeling} \label{cloudy}

As stated in the main text, we vary the metallicity from Solar, $-0.5$ (LMC), $-0.7$ (SMC), to $-1.5$ (I Zw 18) using ``metals and grains -X log''. We vary the depletion factor (F$_*$) from $-2.0$ to $0$ by steps of $0.1$ using ``metals deplete Jenkins2009 Fstar=-X vary log''. We vary the ``hden'' parameter from $10^{0}$ cm$^{-3}$ to $10^{3}$ cm$^{-3}$ with $0.5$ dex steps. We ionize the simulated ISM with radiation from a ``table star tlusty Ostar'' with an effective temperature of $40000$ K, a surface gravity log(g) of 4.0, and a metallicity equal to the host galaxy metallicity. We vary the brightness of the ionizing source by setting the ionization parameter to vary between $\log(\rm U) = 0$ to $\log(\rm U) = -4$ with $0.5$ dex steps. We use ``abundances H II region'' and ``sphere'' as recommended by the documentation. We set the inner radius to $10^{16}$ cm. We set the stopping condition to an ionized hydrogen column density of $10^{20}$ cm$^{-2}$. 

\section{Grain Growth Timescale From H\,\textsc{ii} Region Depletion} \label{timescale}

In the following section, we detail the derivation of the empirical and the theoretical grain growth timescales discussed in Section~\ref{growth}. We begin by estimating the empirical grain growth timescale $\tau_{\rm g,\,observed}$. We denote the initial and final fraction of Fe in dust for the evolving gas parcel respectively as $f_{\rm Fe, init}$ and $f_{\rm Fe, fin}$. Following \citet{dwek_evolution_1998, hirashita_dust--gas_1999, zhukovska_evolution_2008, zhukovska_modeling_2016, choban_galactic_2022}, we set the instantaneous change in the fraction of Fe accreted onto dust as

\begin{equation}
\label{eqn:accretion}
    \dot{f}_{\rm Fe}
    = 
    \frac{f_{\rm Fe}(1 - f_{\rm Fe})}{\tau_{\rm g,\,observed}},
\end{equation}

Because we assume that $f_{\rm Fe, init}$ is the fraction of Fe in dust when the gas parcel was homogeneous with the surrounding diffuse gas, the time for grain growth is the lifetime of the molecular cloud $\tau_{MC}$ beginning from the diffuse ISM and ending at the beginning of star formation. Gas accumulates to become a molecular cloud through ISM interactions on a timescale of 15 Myr \citep{hartmann_rapid_2001, vazquez-semadeni_molecular_2007}. We can additionally assume a molecular cloud collapses and form stars on a timescale of 5 Myr \citep{vazquez-semadeni_molecular_2007, chevance_molecular_2020}, thus $\tau_{MC}\sim20$ Myr. Finally, to find the accretion timescale we solve

\begin{equation}
    f_{\rm Fe, fin} - f_{\rm Fe, init} =\int^{\tau_{\rm MC}}_{0} \dot{f}_{\rm Fe} (t) \, dt.
\end{equation}

Using $f_{\rm Fe, init}$ and $f_{\rm Fe, fin}$ from Table~\ref{tab:hii-abundancess} and Table~\ref{tab:hi-abundances-summary}, we obtain grain growth timescales listed in Table~\ref{tab:grain-growth-timescales}. Note that $\tau_{\rm g}$ is linear with the assumed $\tau_{MC}$. Thus if we assume $\tau_{MC}\sim10$ Myr following \citet{zhukovska_evolution_2008}, the estimated grain growth timescales would be reduced by half.

\begin{table*}[ht!]
\centering
\caption{Estimated Grain Growth Timescales for Each H\,\textsc{ii} Region, $\tau_{\rm MC}=20$ Myr}
\begin{tabular}{ccccccc}
\toprule
30 Doradus & N11B & N66A & N81 & N88A & N90 & \\
\midrule
- & 45.6 $^{+101.9}_{-18.8}$ & 11.7 $^{+1.5}_{-1.2}$ & 12.4 $^{+1.3}_{-1.2}$ & 326 $^{+\infty}_{-218}$ & $<5.9$ $^{+0.4}_{-0.3}$ & Myr \\
\bottomrule
\end{tabular}
\smallskip
\label{tab:grain-growth-timescales}
\end{table*}

\begin{figure*}[ht!]
    \centering
    \includegraphics[width = 4in]{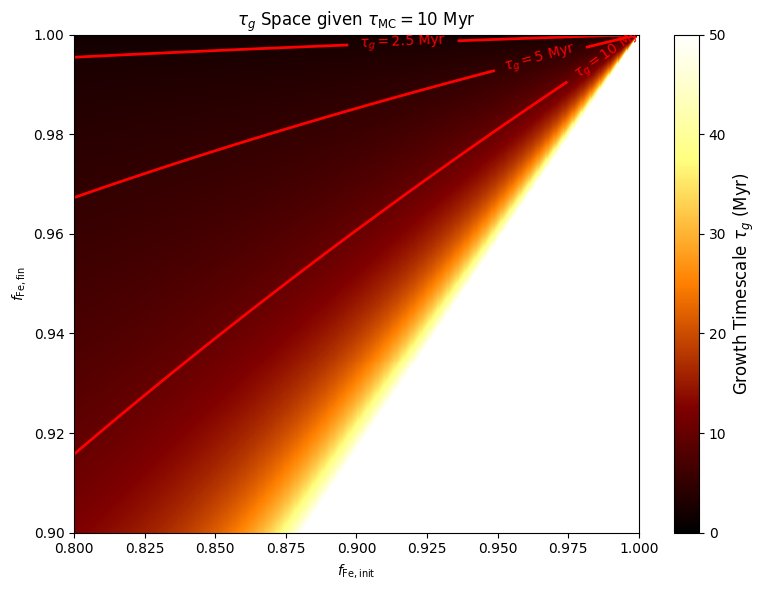}
    \caption{Change of the inferred $\tau_{\rm g,\,observed}$ relative to the initial and final fraction of Fe in dust for an assumed molecular cloud lifetime of 10 Myr. The horizontal axis denotes the initial depleted Fe fraction, which we assume is represented by the Fe depletion measured from the surrounding diffuse ISM; the vertical axis represents the final depleted Fe fraction, where the H\,\textsc{ii} region depletion sets the lower limit. The colorbar shows the growth timescale, with the $\tau_{\rm g,\,observed}\sim$ 2.5, 5, and 10 Myr contours shown in red. }
    \label{fig:tau_growth}
\end{figure*}

The full parameter dependence for $\tau_{\rm g,\,observed}$ is shown in Figure~\ref{fig:tau_growth}. Note that the derived LMC growth timescale is longer (implying slower grain growth). Fe depletion measured from the surrounding diffuse ISM for the LMC H\,\textsc{ii} regions have high N(H). Therefore, these surrounding gas are likely partially molecular themselves and have underwent more grain growth since when they were homogeneous with the gas parcel of interest. 

The expected $\tau_{\rm g}$ for non-Coulomb enhanced Fe-bearing grains can be estimated using the expression from \citet{zhukovska_evolution_2008, zhukovska_modeling_2016, zhukovska_iron_2018, choban_galactic_2022}. For this derivation, we assume all gas-phase Fe are atomic, all Fe-bearing grains are in the form of pure metallic iron \citep[e.g. observationally,][]{mcdonald_rusty_2010, hensley_thermodynamics_2017, boyer_discovery_2025}, and that such grains are spherical: 

\begin{equation}  \label{eqn:growth_timescale}
    \tau_{\rm g,\,theoretical}
    = 
    \frac{\rho_{\rm c} \left<a\right>_{3}}
    {3 \xi_{\rm Fe} \varv_{j,{\rm Fe,th}} A_{\rm Fe} m_{\rm H} \, n_{\rm Fe}},
\end{equation}

where $\rho_{\rm c}=7.86 \; {\rm g\,cm^{-3}}$ is the mass density of metallic iron dust. $\xi_{\rm Fe}$ is the sticking efficiency for each gas-dust collision: at low temperatures, \citet{bossion_accurate_2024} find the sticking efficiency for carbonaceous dust is of order of unity; thus we assume $\xi_{\rm Fe}=1$. $v_{\rm Fe,th}=\sqrt{\frac{8\, k\, T}{\pi A_{\rm Fe}m_{\rm H}}}$ is the thermal velocity of gas-phase Fe with $T=20$ K. $A_{\rm Fe} \, m_{\rm H}$ is the mass of one Fe atom, added to the dust grain with each collision. $n_{\rm Fe}$ is the maximum number density of gas-phase Fe (i.e.\ assuming no depletion onto dust); a typical molecular cloud density ranges from $10^2$ cm$^{-3}$ to up to $10^5$ cm$^{-3}$ in the core \citep{chevance_molecular_2020}, we assume $n(\rm H)=10^3$ cm$^{-3}$ similar to \citet{choban_galactic_2022}; the SMC and LMC photospheric 12 + log(Fe/H) is 6.89 and 7.32 respectively \citep{tchernyshyov_elemental_2015}; thus, $n_{\rm Fe}$ is $7.8\times10^{-3}$ cm$^{-3}$ and $2.1\times10^{-2}$ cm$^{-3}$ respectively. Finally, $\left<a\right>_{3}$ is the average grain radius given by
\begin{equation} \label{eqn:avg_grain_size}
    \left<a\right>_{3} 
    =  
    \frac{\left<a^{3}\right>}{\left<a^{2}\right>} 
    = 
     \frac{\int^{a_{\rm max}}_{a_{\rm min}} \frac{dn_{\rm gr}(a)}{da} \; a^3 \; da}{\int^{a_{\rm max}}_{a_{\rm min}} \frac{dn_{\rm gr}(a)}{da} \; a^2 \; da},
\end{equation}
and $n_{\rm gr}(a)$ is the grain size distribution with minimum and maximum grain sizes $a_{\rm min}$ and $a_{\rm max}$ respectively.
We adopt a MRN size distribution $\frac{dn_{\rm gr}(a)}{da} \propto a^{-3.5}$ \citep{mathis_size_1977} with $a_{\rm min}=4$ nm and $a_{\rm max}=250$ nm for regular-sized metallic iron grains, and $a_{\rm min}=1$ nm and $a_{\rm max}=10$ nm for iron nanoparticles (with $\frac{dn_{\rm gr}(a)}{da} \propto a^{-3.5}$). We obtain $\left<a\right>_{3}=31.6$ nm for regular metallic iron grains and $\left<a\right>_{3}=3.16$ nm for iron nanoparticles. Therefore, Eq.~\ref{eqn:growth_timescale} numerically evaluates to

\begin{align*}
    \tau_{\rm g,\,theoretical} 
    &= 
    41.3 \; {\rm Myr}\text{ - SMC, MRN} \\
    &=
    15.4 \; {\rm Myr}\text{ - LMC, MRN} \\
    &=
    4.18 \; {\rm Myr}\text{ - SMC, nano particles} \\
    &=
    1.56 \; {\rm Myr}\text{ - LMC, nano particles}
\end{align*}

For the diffuse ISM with $n(\rm H)=30$ cm$^{-3}$ and $T=100$ K, the growth timescales are 621 Myr (SMC, MRN), 231 Myr (LMC, MRN), 62.1 Myr (SMC, nano particles), 23.1 Myr (LMC, nano particles). 

For an exploratory analysis, we may use the theoretical diffuse ISM growth timescales to estimate the ``true'' initial depletion. Assuming a total molecular cloud and H\,\textsc{ii} region age of 20 Myr, we integrate Eq.~\ref{eqn:accretion} to obtain the corrected initial fractions of Fe in dust for the diffuse ISM listed in Table~\ref{tab:fraction-of-fe-in-dust}. Using the corrected initial fractions, we may estimate the grain growth timescale again for the observed H\,\textsc{ii} region Fe depletion. The final timescales are listed in Table~\ref{tab:grain-growth-timescales-corrected}. 

\begin{table*}[ht!]
\centering
\caption{Initial Fractions of Fe in Dust, Corrected with Grain Growth in the Diffuse ISM}
\begin{tabular}{cccc}
\toprule
H\,\textsc{ii} region & Uncorrected & Corrected with MRN Fe grains & Corrected with iron nanoparticles\\
\midrule

30 Doradus & 0.976 & 0.974 & 0.945 \\
LMC N11B & 0.980 & 0.978 & 0.954 \\
SMC N66A & 0.885 & 0.882 & 0.848 \\
SMC N81 & 0.830 & 0.825 & 0.780 \\
SMC N88A & 0.947 & 0.945 & 0.928 \\
SMC N90 & 0.850 & 0.846 & 0.804 \\
\bottomrule
\end{tabular}
\smallskip
\label{tab:fraction-of-fe-in-dust}
\end{table*}

\begin{table*}[ht!]
\centering
\caption{Estimated Grain Growth Timescales for Each H\,\textsc{ii} Region with Corrected Initial Depletion, $\tau_{\rm MC}=20$ Myr}
\begin{tabular}{cccc}
\toprule
H\,\textsc{ii} region & Uncorrected & Corrected with MRN Fe grains & Corrected with iron nanoparticles\\
\midrule

30 Doradus & - & $\leq244$ Myr & $\leq23.2$ Myr \\
LMC N11B & 45.6 Myr & 37.4 Myr & 15.4 Myr \\
SMC N66A & 11.7 Myr & 11.5 Myr & 9.9 Myr \\
SMC N81 & 12.4 Myr & 12.1 Myr & 10.3 Myr \\
SMC N88A & 326 Myr & 199 Myr & 51.6 Myr \\
SMC N90 & $<5.9$ Myr & $<5.9$ Myr & $<5.4$ Myr \\
\bottomrule
\end{tabular}
\smallskip
\label{tab:grain-growth-timescales-corrected}
\end{table*}

\end{document}